\documentclass[aps,prl,preprint,superscriptaddress]{revtex4-1}

\usepackage{bm}
\usepackage{graphicx}
\usepackage{amsmath}
\usepackage{hyperref}
\usepackage{epsfig}
\usepackage{epstopdf}

\usepackage[ngerman, english]{babel}

\DeclareGraphicsExtensions{.eps}

\usepackage{hyperref}
\hypersetup{%
	colorlinks=true,
	linkcolor=blue,
	urlcolor=blue,
	citecolor=blue
}
\mathchardef\mhyphen="2D

\begin{document}
	
	
	\title{Effects of processing on the stability of molybdenum oxide ultra-thin films}
	
	
	\author{Aitana Tarazaga Mart\'in-Luengo}
	\email{aitana.tarazaga@jku.at}
	\affiliation{Institut f\"ur Halbleiter-und-Festk\"orperphysik, Johannes Kepler University, Altenbergerstr. 69, A-4040 Linz, Austria}
	\affiliation{Linz Institute of Technology - LIT, Johannes Kepler University, Altenbergerstr. 69, A-4040 Linz, Austria}
	
	\author{Harald K$\mathrm{\ddot{o}}$stenbauer}
	\affiliation{PLANSEE SE, Reutte, Austria}
	
	\author{J$\mathrm{\ddot{o}}$rg Winkler}
	\affiliation{PLANSEE SE, Reutte, Austria}
	
	\author{Alberta Bonanni}
	\email{alberta.bonanni@jku.at}
	\affiliation{Institut f\"ur Halbleiter-und-Festk\"orperphysik, Johannes Kepler University, Altenbergerstr. 69, A-4040 Linz, Austria}
	\affiliation{Linz Institute of Technology - LIT, Johannes Kepler University, Altenbergerstr. 69, A-4040 Linz, Austria}
	
	
	\date{\today}
	
	\begin{abstract}
		
		The effects of wet chemical processing conventionally employed in device fabrication standards are systematically studied on molybdenum oxide ($\mathrm{Mo}\mathrm{O}_{x}$) ultra-thin films. We have combined x-ray photoelectron spectroscopy (XPS), angle resolved XPS and x-ray reflectivity techniques to provide deep insights into the changes in composition, structure and electronic states upon treatment of films with different initial stoichiometry prepared by reactive sputtering. Our results show significant reduction effects associated with the development of gap states in $\mathrm{Mo}\mathrm{O}_{x}$, as well as changes in the composition, density and structure of the films, systematically correlated with the initial oxidation state of Mo.

	\end{abstract}
	
	
	\maketitle
	


	\section{Introduction}
	Molybdenum oxide ($\mathrm{Mo}\mathrm{O}_{x}$) is a transition metal oxide showing extraordinary and versatile electrical, structural, chemical and optical properties, which depend on the oxidation state of Mo, on the degree of crystallinity, on the sample morphology and on environmental conditions. This material system, particularly in the form of thin and ultra-thin films, finds applications in a variety of technologically relevant fields, including catalysis \cite{Brookes:2016_Catalyst,Tagliazucca:2013_J.Catal.}, gas sensors \cite{Illyaskutty_2013:Sensor.Actuat.B-Chem.,Rahmani:2010_Sensor.Actuat.B-Chem.}, optically switchable coatings \cite{Okumu:2006_ThinSolidFilms,AlKuhaili:2015_Mater.Design.}, building-blocks for high-energy density solid-state microbatteries \cite{Ramana:2012_J.Vac.Sci.Technol.A,Rao:2013_Res.J.Recent.Sci.}, smart windows technology \cite{Gesheva:2014_J.Phys.Conf.Ser.,Runnerstrom:2014_Chem.Commun.}, flexible supercapacitors \cite{Cao:2015_Adv.Mater.}, thin film transistors (TFTs) \cite{Chu:2005_Appl.Phys.Lett.} and organic electronics \cite{Meyer:2012_Adv.Mater.,Sun:2011_Adv.Mater.,Battaglia:2014_NanoLett.,Dang:2015_Sci.Rep.,Zhang_2014_Adv.Mater.,Zheng:2015_Adv.Mater.,Elumalai_2015_Mater.Renew.Sustain.Energy,Vasilopoulou:2012_Appl.Phys.Lett.,Meyer:2011_Adv.Mat.,Meyer:2014_Sci.Rep.,Minari:2009_Appl.Phys.Lett.}. Owing to its high work function -- reported to reach 6.9 eV [\onlinecite{Meyer:2012_Adv.Mater.}] -- and to the layered structure of $\alpha\mhyphen\mathrm{Mo}\mathrm{O}_3$, $\mathrm{Mo}\mathrm{O}_{x}$ is currently also employed as a 2D material beyond graphene and as efficient hole contact on 2D transition metal dichalcogenides for \textit{p}-type field effect transistors (\textit{p}-FETs) \cite{Balendhran:2013_Adv.Funct.Mater.,Balendhran:2013_Adv.Mater.,McDonnell:2014_ACSNano}. 
	In view of a reliable device performance, the control over the chemical and physical properties of the $\mathrm{Mo}\mathrm{O}_{x}$ system is mandatory. It has been recently reported \cite{Kostis:2013_J.Phys.Chem.C,Jasieniak:2012_Adv.Funct.Mater.,Murase:2012_Adv.Mater.,Bao:2010_Appl.Phys.Lett.,Battaglia:2014_NanoLett.,Greiner:2011_Nat.Mater.,Wong:2012_J.Phys.Chem.C,Greiner:2012_Adv.Funct.Mater.} that in $\mathrm{Mo}\mathrm{O}_{x}$ with $x<3$, oxygen vacancies originated from partially populated \textit{d}-states, give rise to occupied energy states within the forbidden gap -- reported to be $\sim$ 3.0 eV at room temperature \cite{Meyer:2012_Adv.Mater.} -- becoming bands above a critical concentration and driving the Fermi level close to the conduction band. 
	The oxygen vacancies concentration, and consequently the averaged oxidation state of Mo, is a key parameter directly affecting the properties of the $\mathrm{Mo}\mathrm{O}_{x}$ system. When fully oxidized, $i.e.$ for $\mathrm{Mo}\mathrm{O}_3$ with the corresponding formal oxidation state 6+, $\mathrm{Mo}\mathrm{O}_{x}$ is a standard closed \textit{d}$^{0}$ oxide, transparent and devoided of oxygen vacancies. In this case, there are no occupied states within the wide band gap and the material follows an insulating behaviour. 
	By reducing the metal cations, that is, by increasing the amount of oxygen vacancies, $\mathrm{Mo}\mathrm{O}_3$ forms a series of stable and metastable suboxides ($\mathrm{Mo}\mathrm{O}_{x}$, $2<x<3$), reported $e.g.$ by Magn\'eli \cite{Magneli:1948_Acta.Chem.Scand.} and Kihlborg \cite{Kihlborg:1959_Acta.Chem.Scand.}, showing semiconducting behaviour and a gradually increasing opacity. These sub-stoichiometric oxides present a complex x-ray photoelectron spectroscopy (XPS) spectrum, resulting from the convolution of three different Mo valence states, namely +6, +5 and +4. 
	Moreover, $\mathrm{Mo}\mathrm{O}_2$, with formal oxidation state +4 and with the 4\textit{d}$^{2}$ electron configuration, shows metallic conductivity and the highest opacity of the whole series. Lately, the effects of air exposure on $\mathrm{Mo}\mathrm{O}_{x}$ have attracted considerable attention \cite{Ifran:2012_J.Photon.Energy,Ifran:2012_Appl.Phys.Lett.,Huang:2014_Sci.Rep.}. For instance, recent developments in the field of organic solar cells have shown the critical impact of air exposure on the electronic structure of the $\mathrm{Mo}\mathrm{O}_x$ surface and, consequelty, on the device efficiency and lifetime \cite{Kroger:2009_Appl.Phys.Lett,Ifran_2010_Appl.Phys.Lett.,Vasilopoulou:2012_J.Am.Chem.Soc.,Vasilopoulou_2014_J.Phys.Chem.Lett.,Soultati:2016_J.Phys.Chem.C,Bovill:2013_Appl.Phys.Lett.,Meyer:2010b_Appl.Phys.Lett.,Meyer:2012_Adv.Mater.,Lee:2014_J.Phys.Chem.}. 
	Surface hydration or hydroxylation are suggested as possible mechanisms for reduction of cations in the $\mathrm{Mo}\mathrm{O}_{x}$ system and, thus, for electronic structure changes. In particular, the hydroxylation mechanism is supported by recent density functional theory (DFT) calculations \cite{Butler:2015_Appl.Phys.Lett.}. Moreover, several reports from the field of organic electronics \cite{Kanai:2010_Org.Electron.,Meyer:2012_Adv.Mater.,Kroger:2009_Appl.Phys.Lett,Meyer:2010_Appl.Phys.Lett.} point at gap states formed not only by oxygen vacancies and cation reduction, but also by organic adsorbates at the interface between $\mathrm{Mo}\mathrm{O}_{x}$ and the organic layers.
	
	Process engineering for devices fabrication generally involves standardized steps in which thin layers are treated with various photo-resists and solvents, but systematic studies on the effects of processing on the properties of $\mathrm{Mo}\mathrm{O}_{x}$ films are still wanted. In this work, we report on the effects of processing on the stability of ultra-thin $\mathrm{Mo}\mathrm{O}_{x}$ layers treated with conventional photo-resist (Shipley S1818) and solvents. The in-depth study is carried out on a series of $\sim$ 10 nm thick amorphous $\mathrm{Mo}\mathrm{O}_{x}$ layers fabricated by means of reactive sputtering and presenting different stoichiometry, spanning -- over the series -- from a metallic to a fully oxidized phase and including, in particular, Mo, $\mathrm{Mo}\mathrm{O}_{2+\epsilon}$, $\mathrm{Mo}\mathrm{O}_{3-\epsilon}$, and $\mathrm{Mo}\mathrm{O}_3$.

	\section{Experimental Procedure}
	A combination of XPS, angle resolved XPS (ARXPS) and x-ray reflectivity (XRR) techniques is employed to determine the charge state of the cations, the valence band spectra and the structural arrangement of the samples as-grown and upon treatment/exposure. A list of the as-grown samples under study is provided in Table\,II of the Appendix, while the labeling followed to identify the samples series according to the exposure/processing, is given in Table\,I. 
	
		\begin{table}[h]
			\centering
			\begin{tabular}{|c|l|}
				\hline
				A$_\mathrm{v}$ & Virgin samples. Shortly exposed to air after growth. \\ \hline
				A$_\mathrm{ref}$ &  \begin{tabular}[c]{@{}l@{}}Reference samples. Stored under cleanroom condition \\ until stabilization and used for processing.\end{tabular} \\ \hline
				B$_\mathrm{0}$ & \begin{tabular}[c]{@{}l@{}}Samples measured 1h after exposure to conventional \\ cleaning protocol by ultrasonic treatment in acetone, \\ 2-propanol and DI water sequentially.\end{tabular} \\ \hline
				B$_\mathrm{1}$ & \begin{tabular}[c]{@{}l@{}}Samples measured 4 days after esposure to conventional \\ cleaning protocol by ultrasonic tratment in acetone, \\ 2-propanol and DI water sequentially.\end{tabular} \\ \hline
				C$_\mathrm{0}$ & \begin{tabular}[c]{@{}l@{}}Samples measured 1h after exposure to conventional \\ cleaning protocol by ultrasonic treatment in acetone, \\ 2-propanol sequentially.\end{tabular} \\ \hline
				C$_\mathrm{1}$ & \begin{tabular}[c]{@{}l@{}}Samples measured 4 days after exposure to conventional \\ cleaning protocol by ultrasonic treatment in acetone, \\ 2-propanol sequentially.\end{tabular} \\ \hline
			\end{tabular}
			\caption{Labeling of the samples series according to the exposure/processing.}
			\label{tab:tab1}
		\end{table}


	The as-grown virgin samples, labeled as $\mathrm{A}_\mathrm{v}$, are stored under cleanroom conditions until they are stabilized after chemisorption of oxygen and moisture on the surface. Once stabilized, that is, once the samples show stable XPS spectra measured a week apart, these reference samples $\mathrm{A}_\mathrm{ref}$ are coated with photo-resist which is then removed following the different standard procedures considered here below and summarized in Table\,I. When treated according to a conventional cleaning protocol of 15 minutes in acetone ultrasonic bath followed by 15 minutes in 2-propanol, and ending with 1 minute rinse in DI water, the previously stabilized $\mathrm{A}_\mathrm{ref}$ samples are referred to as samples B. When subjected to the previous procedure, but avoiding the final rinse in DI water, the samples are labelled as series C. Subscripts 0 and 1 ($\mathrm{B}_{0}$, $\mathrm{B}_{1}$, $\mathrm{C}_{0}$, and $\mathrm{C}_{1}$) correspond to samples measured one hour and four days after processing, respectively.
	
	The series of amorphous $\mathrm{Mo}\mathrm{O}_{x}$ ultra-thin films are reactively sputtered in DC mode on glass substrates (Corning Eagle XG) at room temperature. The high resolution XPS and ARXPS spectra, with probing depth of 5-10 nm, are collected using Al $K$$\alpha$ radiation (1486.6 eV), with a constant pass energy of 50 eV (1.00 eV full-width-at-half-maximum on Ag $\mathrm{3\textit{d}}_\mathrm{5/2}$) and energy step size of 0.05 eV. The binding energies (BEs) are calibrated with respect to the conventional C 1$s$ peak at 284.8 eV. For the XRR measurements, a wavelength of 1.54 $\textrm{\AA}$ in a Seifert XRD 3003 PTS-HR is employed and the GenX reflectivity fitting package \cite{Bjorck:2007_J.Appl.Crystallogr.} supports the data analysis. Details on growth parameters and XRR measurements are provided in the Appendix.

	\section{Results and Discussion}
	
\begin{figure}[ht]
		\centering
		\includegraphics[width=6 cm]{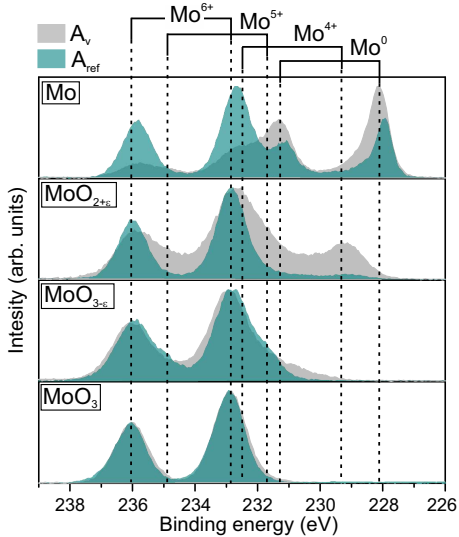}
		\caption{Comparison between high resolution Mo 3\textit{d} XPS spectra for the samples as-grown $\mathrm{A}_\mathrm{v}$ and upon stabilization in cleanroom atmosphere $\mathrm{A}_\mathrm{ref}$.}
		\label{fig:fig1}
	\end{figure}

	In Fig.\,\ref{fig:fig1} the high resolution Mo 3\textit{d} core level XPS spectra for the samples: (i) $\mathrm{A}_\mathrm{v}$ as-grown and (ii) $\mathrm{A}_\mathrm{ref}$ upon stabilization in cleanroom conditions are reported. For each valence state, the core level is split into the $\mathrm{3\textit{d}}_{5/2}$ and $\mathrm{3\textit{d}}_{3/2}$ doublets, separated by 3.1 eV, and with binding energies assigned according to established reference values \cite{Baltrusaitis:2015_Appl.Surf.Sci.}. A significant surface oxidation is observed for all the samples exposed to atmosphere. As expected, the oxidation effect is more pronounced in virgin samples with a more metallic character. 
	
	\begin{figure*}[ht]
		\centering
		\includegraphics[width=17 cm]{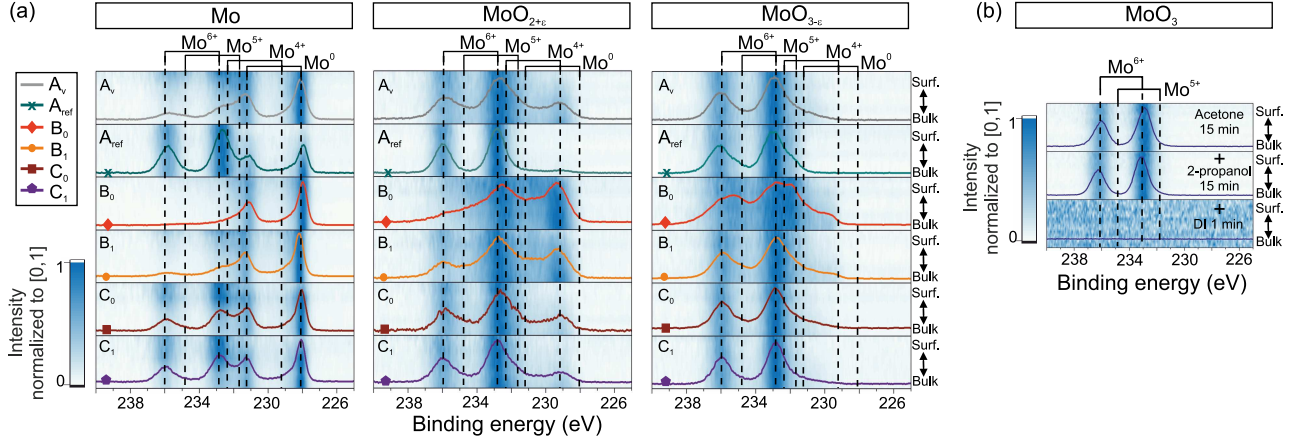}
		\caption {High resolution Mo 3\textit{d} ARXPS intensity maps as a function of depth and overall XPS intensity for (a) Mo, $\mathrm{Mo}\mathrm{O}_{2+\epsilon}$ and $\mathrm{Mo}\mathrm{O}_{3-\epsilon}$, and (b) $\mathrm{Mo}\mathrm{O}_{3}$ before and after processing.}  
		\label{fig:fig2} 	
	\end{figure*}

	Detailed ARXPS depth profiles for the virgin, reference (upon stabilization in cleanroom atmosphere), and processed B and C series, are represented in Fig.\,\ref{fig:fig2}. As evidenced in the left panel of Fig.\,\ref{fig:fig2}(a), upon stablization in cleanroom atmosphere, the metallic component $\mathrm{Mo}^\mathrm{0}$ of the pure Mo film vanishes from the first atomic layers, while the fully oxidized valence state $\mathrm{Mo}^\mathrm{6+}$ becomes predominant over the whole profile, as expected from oxidation of a metal. 
	On the other hand, for $\mathrm{Mo}\mathrm{O}_{2+\epsilon}$, and as reported in the middle panel of Fig.\,\ref{fig:fig2}(a), $\mathrm{Mo}^\mathrm{4+}$ and $\mathrm{Mo}^\mathrm{5+}$ are significantly quenched upon exposure to atmosphere, while the intensity of the $\mathrm{Mo}^\mathrm{5+}$ component decreases for the two oxidation states with more oxygen content, $\mathrm{Mo}\mathrm{O}_{3-\epsilon}$ and $\mathrm{Mo}\mathrm{O}_{3}$, but it is not completely suppressed, as evidenced in the right panel of Fig.\,\ref{fig:fig2}(a). 
	
	For all the initial stoichiometries, the oxidation process is not confined to the uppermost atomic layers, but interests the whole probing depth. 
	
	The reduction effect reported previously \cite{Meyer:2012_Adv.Mater.,Ifran_2010_Appl.Phys.Lett.} and due to air exposure is not found in this work, likely due the fact that the samples considered in literature were not completely stabilized, while we have ensured and confirmed by XPS analysis, that the $\mathrm{A}_\mathrm{ref}$ films studied here are fully stabile. In fact, it was theoretically demonstrated by Butler $et$ $al$. \cite{Butler:2015_Appl.Phys.Lett.}, that essential properties of the $\mathrm{Mo}\mathrm{O}_{x}$ system, like $e.g.$ the ionization potential, critically depend on the exposure to moisture until stabilization. 
	
	By considering the samples $\mathrm{B}_{0}$, $i.e.$ the $\mathrm{A}_\mathrm{ref}$ upon conventional treatment involving DI water, it is inferred from Fig.\,\ref{fig:fig2}(a) that in the sub-stoichiometric samples $\mathrm{Mo}\mathrm{O}_{2+\epsilon}$ and $\mathrm{Mo}\mathrm{O}_{3-\epsilon}$, the Mo 3\textit{d} emission does not diminish significantly, but shifts to lower BE, pointing to a significant chemical reduction enhanced for the bulk angle, that is, for deep atomic layers. After four days of atmospheric exposure, $i.e.$ $\mathrm{B}_{0}$$\rightarrow$$\mathrm{B}_{1}$, further oxidation leads to a lowering of the intensity of the $\mathrm{Mo}^\mathrm{4+}$ emission, while the one of $\mathrm{Mo}^\mathrm{6+}$ is augmented, recovering a spectrum resembling the one obtained from the virgin samples $\mathrm{A}_\mathrm{v}$ before stabilization. 
	
	Upon processing without DI water, $i.e.$ $\mathrm{A}_\mathrm{ref}$$\rightarrow$$\mathrm{C}_{0}$, in both sub-stoichiometric compounds $\mathrm{Mo}\mathrm{O}_{2+\epsilon}$ and $\mathrm{Mo}\mathrm{O}_{3-\epsilon}$ the reduction effect is less pronounced than upon processing with DI water. The features of the virgin samples $\mathrm{A}_\mathrm{v}$ one hour after treatment ($\mathrm{C}_{0}$) are recovered, and no remarkable changes are detected after four days of atmospheric exposure ($\mathrm{C}_{1}$). In the case of the pure metallic layer, processing with DI water ($\mathrm{A}_\mathrm{ref}$$\rightarrow$$\mathrm{B}_{0}$) fosters the full removal of the superficial oxide layer, while upon treatment without DI water, the oxide layer is reduced in thickness. The particular case of the fully oxidized sample after processing, $\mathrm{Mo}\mathrm{O}_{3}$, is shown in Fig.\,\ref{fig:fig2}(b). No remarkable changes are detectable after each phase of the cleaning protocol, until the last step with DI water is applied. Only upon water exposure the emission from the Mo 3\textit{d} core level is completely quenched, pointing to a complete removal of Mo from the layer. 
		
\begin{figure}[ht]
		\centering
		\includegraphics[width=8.5 cm]{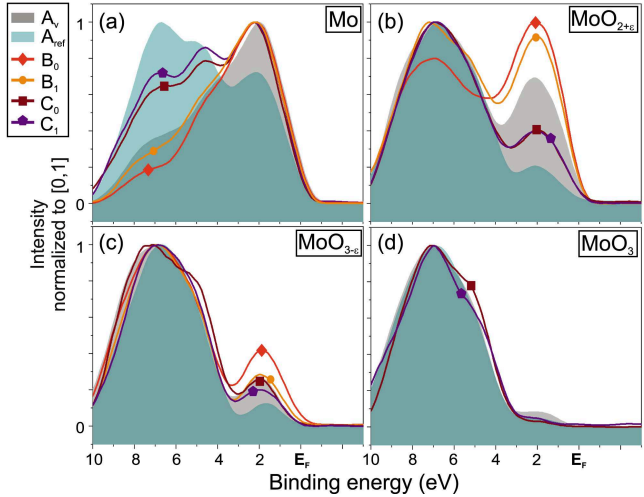}
		\caption{Normalized evolution of the valence band region before and after processing for (a) Mo, (b) $\mathrm{Mo}\mathrm{O}_{2+\epsilon}$, (c) $\mathrm{Mo}\mathrm{O}_{3-\epsilon}$, and (d) $\mathrm{Mo}\mathrm{O}_{3}$.}  
		\label{fig:fig3}
	\end{figure}

	The response of the Mo layers summarized in the left panel of Fig.\,\ref{fig:fig2}(a) is confirmed by the evolution of the XPS valence band energy distribution curves of Mo depicted in Fig.\,\ref{fig:fig3}(a), and characterized by a broad overall band with a maximum intensity at 2.0 eV below the Fermi level \cite{Werfel:1983_J.Phys.C.SolidStatePhys.} $\mathrm{E}_\mathrm{F}$. The broad emission centered at $\sim 6.0$ eV is assigned to an O 2$p$ photoemission signal of adsorbed oxygen, with contributions from various oxygen species. Its intensity is correlated with the presence of $\mathrm{Mo}^\mathrm{6+}$ and $\mathrm{Mo}^\mathrm{5+}$ oxidation states in the Mo 3\textit{d} core level spectra in Fig.\,\ref{fig:fig2}(a). As evidenced in Figs.\,\ref{fig:fig3}(b-d), in the case of the oxide films, the valence band region shows the conventional transition metal oxide two band structure \cite{Greiner:2011_Nat.Mater.} resulting in: (i) one peak mainly due to O 2$p$ orbitals centered at $\sim$6 eV, and which identifies the valence band maximum (VBM) at $\sim$3 eV, (ii) and a second peak emerging between the VBM and the $\mathrm{E}_\mathrm{F}$. In the case of the intermediate oxides, this latter band of gap states originates from oxygen vacancies partially filling the empty Mo $d$ levels. 
	
	For the as-grown samples before processing, there is a systematic correlation between the Mo oxidation state as in Fig.\,\ref{fig:fig2} and the relative intensity of the emission related to the gap states and of the one from the O 2$p$ levels, as evidenced in Figs.\,\ref{fig:fig3}(a-d). The width of the emission from the gap states is assigned to the presence of different types of vacancies \cite{Papadopoulos:2013_Adv.Funct.Mater.}, whose complex geometry is beyond the resolution of the XPS system. 
	
	Processing involving DI water promotes substantially the development of gap states. This effect is paticularly pronounced for the less oxidized samples $\mathrm{Mo}\mathrm{O}_{2+\epsilon}$, as summarized in Fig.\,\ref{fig:fig3}(b). When $\mathrm{Mo}\mathrm{O}_{2+\epsilon}$ is treated in the absence of DI water ($\mathrm{A}_\mathrm{ref}$$\rightarrow$$\mathrm{C}_{0}$), the intensity of the peak related to the gap states assumes a value intermediate between the one for $\mathrm{A}_\mathrm{v}$ and the one for $\mathrm{A}_\mathrm{ref}$, and it keeps stable ($\mathrm{C}_{0}$$\rightarrow$$\mathrm{C}_{1}$). 	
	As shown in Fig.\,\ref{fig:fig3}(d), for the fully oxidized $\mathrm{Mo}\mathrm{O}_{3}$, the intensity of the initial band states is minimized upon stabilization ($\mathrm{A}_\mathrm{v}$$\rightarrow$$\mathrm{A}_\mathrm{v}$) and does not significantly change after processing without DI water ($\mathrm{A}_\mathrm{ref}$$\rightarrow$$\mathrm{C}_{0},\mathrm{C}_{1}$).

	In all cases, the correlation between the reduction process emerging from the data shown in Fig.\,\ref{fig:fig2} and the evolution of the gap states is preserved, hence we attribute the enhancement of the gap state density to the presence of $\mathrm{Mo}\mathrm{O}_{x}$ reduced states, in accordance with previous reports by other authors \cite{Papadopoulos:2013_Adv.Funct.Mater.,Meyer:2012_Adv.Mater.,Ifran_2010_Appl.Phys.Lett.,Gwinner:2011_Adv.Funct.Mater.}. Moreover, we can state that not only surface states are affected, but also deeper layers within the material, as deduced from the ARXPS measurements in Fig.\,\ref{fig:fig2}.
	
	\begin{figure}[ht]
		\centering
		\includegraphics[width=8.5 cm]{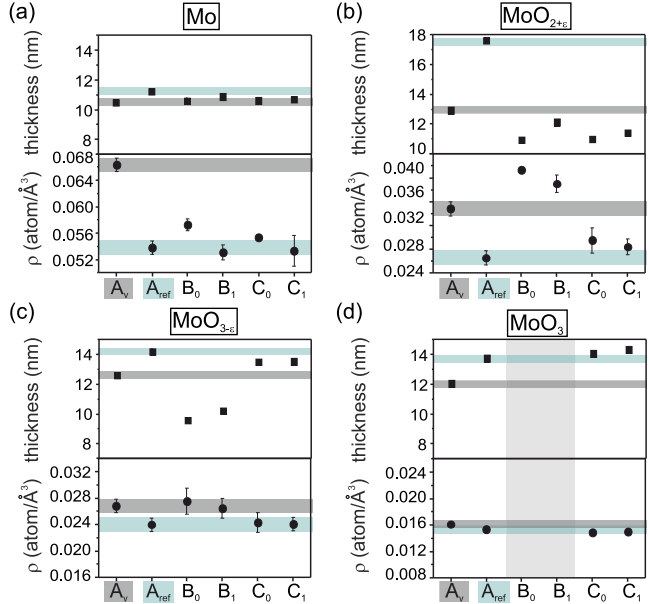}
		\caption{Film density and thickness as deduced from XRR measurements for: (a) Mo, (b) $\mathrm{Mo}\mathrm{O}_{2+\epsilon}$, (c) $\mathrm{Mo}\mathrm{O}_{3-\epsilon}$, and (d) $\mathrm{Mo}\mathrm{O}_{3}$ as-grown, stabilized in cleanroom atmosphere and upon processing. The solid bands are guide for the eyes and mark the values for the as-grown ($\mathrm{A}_\mathrm{v}$) and stabilized ($\mathrm{A}_\mathrm{ref}$) samples.}
		\label{fig:fig4}
	\end{figure}
	
	From XRR measurements -- whose details are provided in the Appendix --  the variation of density and thickness due to exposure/processing has been obtained for the investigated $\mathrm{Mo}\mathrm{O}_\mathrm{x}$ layers and it is summarized in the panels of Fig.\,\ref{fig:fig4}. Generally, an increment of density is reflected in a lowering of the thickness and $vice\,versa$, as inferred from Figs.\,\ref{fig:fig4}(a,c,d). In the case of the $\mathrm{Mo}\mathrm{O}_{2+\epsilon}$ layers of Fig.\,\ref{fig:fig4}(b) after processing in the absence of DI water ($\mathrm{A}_\mathrm{ref}$$\rightarrow$$\mathrm{C}_{0}$), the density remains close to the one of the reference sample, while the thickness is reduced from 17.5\,nm to 11\,nm. Further analysis is required, in order to understand this anomaly. The $\mathrm{Mo}\mathrm{O}_{3}$ films of Fig.\,\ref{fig:fig4}(d), while being the most stable -- as confirmed by XPS measurements -- upon stabilization in cleanroom atmosphere and upon treatment with processes non including DI water, are completely dissociated after 1 min. of exposure to DI water. An intermediate case, is represented by the sub-stoichiometric samples $\mathrm{Mo}\mathrm{O}_{3-\epsilon}$ in Fig.\,\ref{fig:fig4}(c), which -- upon processing with DI water ($\mathrm{A}_\mathrm{ref}$$\rightarrow$$\mathrm{B}_{0}$) -- looses significantly in thickness, pointing at an enhanced solubility of the $\mathrm{Mo}^\mathrm{6+}$ phase in water.
	
	In summary,  reactively sputtered $\mathrm{Mo}\mathrm{O}_{x}$ ultra-thin film layers with different oxidation states (Mo, $\mathrm{Mo}\mathrm{O}_{2+\epsilon}$, $\mathrm{Mo}\mathrm{O}_{3-\epsilon}$, and  $\mathrm{Mo}\mathrm{O}_{3}$) have been exposed to different conventional cleaning protocols and investigated by XPS, ARXPS and XRR. Water strongly reduces the Mo oxidation state for all the samples, and causes the complete dissolution of $\mathrm{Mo}\mathrm{O}_{3}$ after 1 min. of exposure. The high etching selectivity between Mo and $\mathrm{Mo}\mathrm{O}_{3}$ opens wide perspectives for structuring of $\mathrm{Mo}\mathrm{O}_{x}$ by means of simple DI water chemical etching \cite{Rolandi:2002_Adv.Mater.,Espinosa:2015_Appl.Phys.Lett.}. The reduction effect, not confined only to the surface layers, is correlated with the development of gap states -- likely due to chemisorbed species -- which lead to an electron transfer to the transition metal oxide system \cite{Meyer:2012_Adv.Mater.,Ifran_2010_Appl.Phys.Lett.}. Independently of the processing protocol, reduction and gap state intensities are stable after four days of air exposure. We have found that fully metallic and fully oxidized phases have the highest stability against the various cleaning protocols. Our findings indicate that the optimization of the protocols for processing $\mathrm{Mo}\mathrm{O}_{x}$ ultra-thin films are crucial for the engineering of band states, fundamental for $e.g.$ charge transport applications \cite{Kanai:2010_Org.Electron.,Hancox:2010_Org.Electron.}.

\section{Acknowledgements}

This work was supported by the $\mathrm{\ddot{O}}$sterreichische Forschungsf$\mathrm{\ddot{o}}$rderungsgesellschaft FFG within the competence headquarter project $\mathrm{"E}^{2}\mathrm{SPUTTERTECH"}$ (Project 841482), by the Austrian Science Foundation --
FWF (P24471 and P26830), by the NATO Science for Peace Programme (Project 984735), and by the FunDMS Advanced Grant of the European Research Council (ERC grant 227690). The authors acknowledge Jiri Duchoslav and Roland Steinberger for support in the XPS and ARXPS measurements.

\section{Appendix}

\subsection{Growth of $\mathrm{Mo}\mathrm{O}_{x}$ ultra-thin layers}

The $\mathrm{Mo}\mathrm{O}_{x}$ ultra-thin films are reactively sputtered in DC mode from a 100 mm diameter metallic Mo target in Ar/O$_{2}$ atmosphere on glass substrates (Corning Eagle XG) at room temperature. For all the samples the Ar flow rate and the target power are kept constant at 22 sccm and 400 W, respectively. The oxygen partial pressure varies from 0\, mbar, for the pure Mo sample, to 3x10$^{-3}$\,mbar for $\mathrm{Mo}\mathrm{O}_{3}$, controlled by a lambda probe (Zirox vacuum probe). The base pressure is kept at 1x10$^{-6}$\,mbar. The fundamental growth parameters for the different $\mathrm{Mo}\mathrm{O}_{x}$ stoichiometries considered in this work, is provided in Table\,II.

\begin{table*}[h]
	\centering
	\begin{tabular}{cccc|c|c|c|c|c|}
		\cline{5-9}
		&  &  &  & \multicolumn{5}{c|}{Power supply} \\ \hline
		\multicolumn{1}{|c|}{Sample id.} & \multicolumn{1}{c|}{Target} & \multicolumn{1}{c|}{\begin{tabular}[c]{@{}c@{}}p(Ar)\\ (mbar)\end{tabular}} & \begin{tabular}[c]{@{}c@{}}p(O$_{2}$)\\ (mbar)\end{tabular} & \begin{tabular}[c]{@{}c@{}}P\\ (W)\end{tabular} & \begin{tabular}[c]{@{}c@{}}Pdens\\ (W/cm$^{2}$)\end{tabular} & \begin{tabular}[c]{@{}c@{}}U\\ (V)\end{tabular} & \begin{tabular}[c]{@{}c@{}}I\\ (A)\end{tabular} & \begin{tabular}[c]{@{}c@{}}Z\\ (ohms)\end{tabular} \\ \hline
		\multicolumn{1}{|c|}{Mo} & \multicolumn{1}{c|}{Mo} & \multicolumn{1}{c|}{5.0$\times$10$^{-3}$} & - & 400 & 5.1 & 326 & 1.23 & 265 \\ \hline
		\multicolumn{1}{|c|}{\textbf{$\mathrm{Mo}\mathrm{O}_{2+x}$}} & \multicolumn{1}{c|}{Mo} & \multicolumn{1}{c|}{5.0$\times$10$^{-3}$} & 2.5$\times$10$^{-4}$ & 400 & 5.1 & 453 & 0.88 & 515 \\ \hline
		\multicolumn{1}{|c|}{\textbf{$\mathrm{Mo}\mathrm{O}_{3-x}$}} & \multicolumn{1}{c|}{Mo} & \multicolumn{1}{c|}{5.0$\times$10$^{-3}$} & 5.5$\times$10$^{-4}$ & 400 & 5.1 & 553 & 0.72 & 768 \\ \hline
		\multicolumn{1}{|c|}{\textbf{$\mathrm{Mo}\mathrm{O}_{3}$}} & \multicolumn{1}{c|}{Mo} & \multicolumn{1}{c|}{5.0$\times$10$^{-3}$} & 3.0$\times$10$^{-3}$ & 400 & 5.1 & 529 & 0.76 & 696 \\ \hline
	\end{tabular}
	\caption{Growth parameters for the $\mathrm{Mo}\mathrm{O}_{x}$ film stoichiometries under study.}
	\label{tab:tab2}
\end{table*}

\subsection{X-ray reflectivity}

X-ray reflectivity (XRR) data are analyzed using the GenX reflectivity fitting package \cite{Bjorck:2007_J.Appl.Crystallogr.}, where the absolute logarithmic error function is used as a figure of merit (FOM), and it exploits a differential evolution algorithm and the Parratt recursion formula.
A basic two layered structure is defined, consisting of one alkaline earth boro-aluminosilicate substrate and the $\mathrm{Mo}\mathrm{O}_{x}$ layer. The best FOM is generally achieved by adding a second layer of $\mathrm{Mo}\mathrm{O}_{x}$, while keeping the total thickness constant and using the density of the two $\mathrm{Mo}\mathrm{O}_{x}$ layers as parameter. In thist way a gradient of densities due to different oxidation states and/or chemisorbed impurities along the whole profile is simulated. The average density is then employed for the data analysis. The GenX software varies the thickness, density and roughness of each layer and minimizes the difference between model an experimental data. The substrate properties are kept constant, except for the density after processing, which is let free to change within an interval of $\pm$0.01 atoms/$\mathrm{\AA}$, in order to take into account possible changes due to processing. The accuracy achieved for the layer densities and thickness are in all cases $\leq$$\pm$0.002 atoms/$\mathrm{\AA}$ and $\pm$1 nm, respectively. In Fig.\,\ref{fig:fig1}, the experimental and simulated XRR data are shown for all the samples under study before and after processing. 

\begin{figure*}[h]
	\centering
	\includegraphics[width=13 cm]{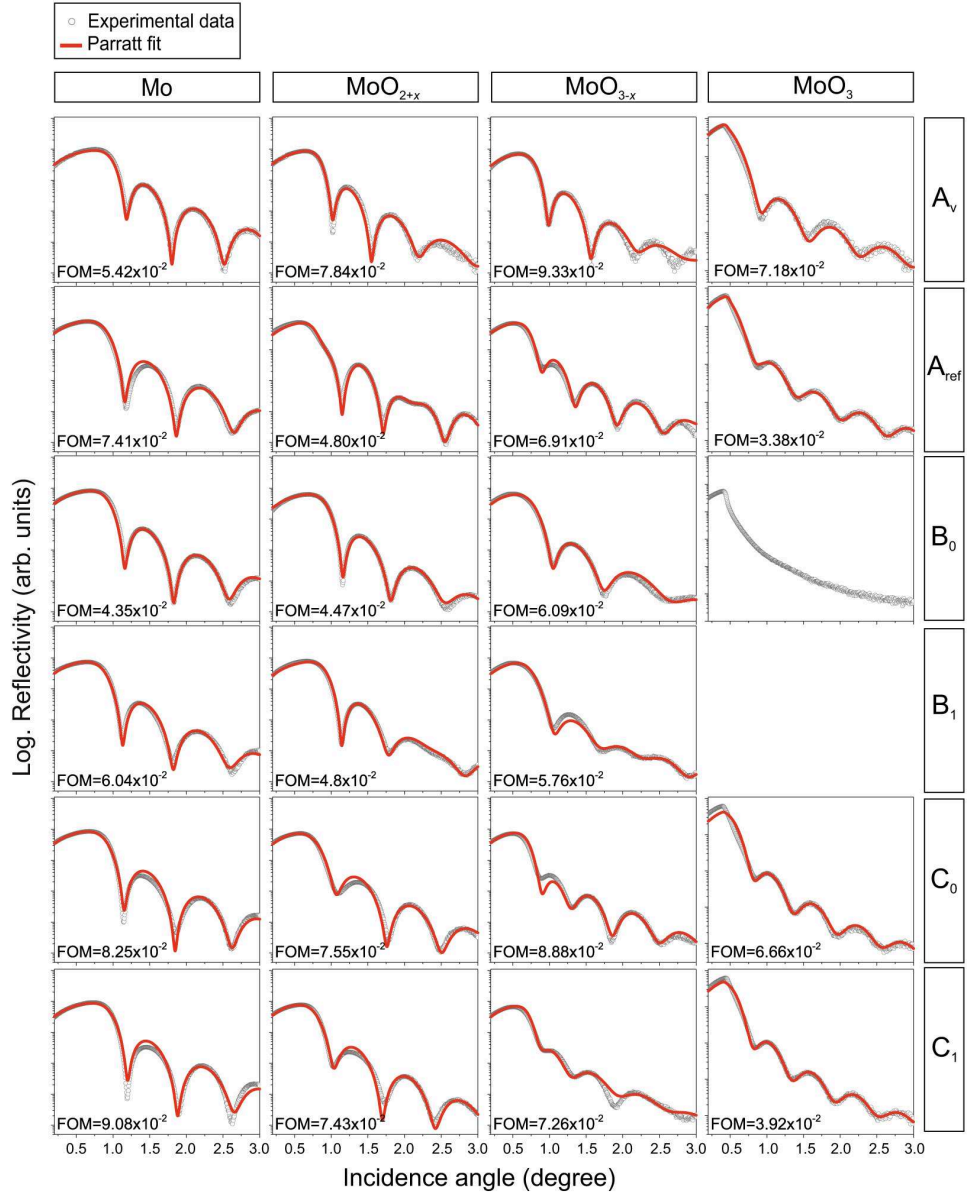}
	\caption{Experimental data and fit of the XRR as a function of the grazing incident angle before and after processing for all the samples under study, $e.i.$, Mo, $\mathrm{Mo}\mathrm{O}_{2+x}$,  $\mathrm{Mo}\mathrm{O}_{3-x}$, and $\mathrm{Mo}\mathrm{O}_{3}$.} 
	\label{fig:fig5}
\end{figure*}

%



\clearpage


\begin{thebibliography}{57}%
	\makeatletter
	\providecommand \@ifxundefined [1]{%
		\@ifx{#1\undefined}
	}%
	\providecommand \@ifnum [1]{%
		\ifnum #1\expandafter \@firstoftwo
		\else \expandafter \@secondoftwo
		\fi
	}%
	\providecommand \@ifx [1]{%
		\ifx #1\expandafter \@firstoftwo
		\else \expandafter \@secondoftwo
		\fi
	}%
	\providecommand \natexlab [1]{#1}%
	\providecommand \enquote  [1]{``#1''}%
	\providecommand \bibnamefont  [1]{#1}%
	\providecommand \bibfnamefont [1]{#1}%
	\providecommand \citenamefont [1]{#1}%
	\providecommand \href@noop [0]{\@secondoftwo}%
	\providecommand \href [0]{\begingroup \@sanitize@url \@href}%
	\providecommand \@href[1]{\@@startlink{#1}\@@href}%
	\providecommand \@@href[1]{\endgroup#1\@@endlink}%
	\providecommand \@sanitize@url [0]{\catcode `\\12\catcode `\$12\catcode
		`\&12\catcode `\#12\catcode `\^12\catcode `\_12\catcode `\%12\relax}%
	\providecommand \@@startlink[1]{}%
	\providecommand \@@endlink[0]{}%
	\providecommand \url  [0]{\begingroup\@sanitize@url \@url }%
	\providecommand \@url [1]{\endgroup\@href {#1}{\urlprefix }}%
	\providecommand \urlprefix  [0]{URL }%
	\providecommand \Eprint [0]{\href }%
	\providecommand \doibase [0]{http://dx.doi.org/}%
	\providecommand \selectlanguage [0]{\@gobble}%
	\providecommand \bibinfo  [0]{\@secondoftwo}%
	\providecommand \bibfield  [0]{\@secondoftwo}%
	\providecommand \translation [1]{[#1]}%
	\providecommand \BibitemOpen [0]{}%
	\providecommand \bibitemStop [0]{}%
	\providecommand \bibitemNoStop [0]{.\EOS\space}%
	\providecommand \EOS [0]{\spacefactor3000\relax}%
	\providecommand \BibitemShut  [1]{\csname bibitem#1\endcsname}%
	\let\auto@bib@innerbib\@empty
	\bibitem [{\citenamefont {Brookes}\ \emph {et~al.}(2016)\citenamefont
		{Brookes}, \citenamefont {Bowker},\ and\ \citenamefont
		{Wells}}]{Brookes:2016_Catalyst}%
	\BibitemOpen
	\bibfield  {author} {\bibinfo {author} {\bibfnamefont {C.}~\bibnamefont
			{Brookes}}, \bibinfo {author} {\bibfnamefont {M.}~\bibnamefont {Bowker}}, \
		and\ \bibinfo {author} {\bibfnamefont {P.~P.}\ \bibnamefont {Wells}},\ }\href
	{\doibase 10.3390/catal6070092} {\bibfield  {journal} {\bibinfo  {journal}
			{Catalysts}\ }\textbf {\bibinfo {volume} {6}},\ \bibinfo {pages} {92}
		(\bibinfo {year} {2016})}\BibitemShut {NoStop}%
	\bibitem [{\citenamefont {Tagliazucca}\ \emph {et~al.}(2013)\citenamefont
		{Tagliazucca}, \citenamefont {Schlichte}, \citenamefont {Sch{\"{u}}th},\ and\
		\citenamefont {Weidenthaler}}]{Tagliazucca:2013_J.Catal.}%
	\BibitemOpen
	\bibfield  {author} {\bibinfo {author} {\bibfnamefont {V.}~\bibnamefont
			{Tagliazucca}}, \bibinfo {author} {\bibfnamefont {K.}~\bibnamefont
			{Schlichte}}, \bibinfo {author} {\bibfnamefont {F.}~\bibnamefont
			{Sch{\"{u}}th}}, \ and\ \bibinfo {author} {\bibfnamefont {C.}~\bibnamefont
			{Weidenthaler}},\ }\href {\doibase 10.1016/j.jcat.2013.05.011} {\bibfield
		{journal} {\bibinfo  {journal} {J. Catal.}\ }\textbf {\bibinfo {volume}
			{305}},\ \bibinfo {pages} {277} (\bibinfo {year} {2013})}\BibitemShut
	{NoStop}%
	\bibitem [{\citenamefont {Illyaskutty}\ \emph {et~al.}(2013)\citenamefont
		{Illyaskutty}, \citenamefont {Kohler}, \citenamefont {Trautmann},
		\citenamefont {Schwotzer},\ and\ \citenamefont {{Mahadevan
				Pillai}}}]{Illyaskutty_2013:Sensor.Actuat.B-Chem.}%
	\BibitemOpen
	\bibfield  {author} {\bibinfo {author} {\bibfnamefont {N.}~\bibnamefont
			{Illyaskutty}}, \bibinfo {author} {\bibfnamefont {H.}~\bibnamefont {Kohler}},
		\bibinfo {author} {\bibfnamefont {T.}~\bibnamefont {Trautmann}}, \bibinfo
		{author} {\bibfnamefont {M.}~\bibnamefont {Schwotzer}}, \ and\ \bibinfo
		{author} {\bibfnamefont {V.~P.}\ \bibnamefont {{Mahadevan Pillai}}},\ }\href
	{\doibase 10.1016/j.snb.2013.05.092} {\bibfield  {journal} {\bibinfo
			{journal} {Sensors Actuators, B Chem.}\ }\textbf {\bibinfo {volume} {187}},\
		\bibinfo {pages} {611} (\bibinfo {year} {2013})}\BibitemShut {NoStop}%
	\bibitem [{\citenamefont {Rahmani}\ \emph {et~al.}(2010)\citenamefont
		{Rahmani}, \citenamefont {Keshmiri}, \citenamefont {Yu}, \citenamefont
		{Sadek}, \citenamefont {Al-Mashat}, \citenamefont {Moafi}, \citenamefont
		{Latham}, \citenamefont {Li}, \citenamefont {Wlodarski},\ and\ \citenamefont
		{Kalantar-zadeh}}]{Rahmani:2010_Sensor.Actuat.B-Chem.}%
	\BibitemOpen
	\bibfield  {author} {\bibinfo {author} {\bibfnamefont {M.}~\bibnamefont
			{Rahmani}}, \bibinfo {author} {\bibfnamefont {S.}~\bibnamefont {Keshmiri}},
		\bibinfo {author} {\bibfnamefont {J.}~\bibnamefont {Yu}}, \bibinfo {author}
		{\bibfnamefont {A.}~\bibnamefont {Sadek}}, \bibinfo {author} {\bibfnamefont
			{L.}~\bibnamefont {Al-Mashat}}, \bibinfo {author} {\bibfnamefont
			{A.}~\bibnamefont {Moafi}}, \bibinfo {author} {\bibfnamefont
			{K.}~\bibnamefont {Latham}}, \bibinfo {author} {\bibfnamefont
			{Y.}~\bibnamefont {Li}}, \bibinfo {author} {\bibfnamefont {W.}~\bibnamefont
			{Wlodarski}}, \ and\ \bibinfo {author} {\bibfnamefont {K.}~\bibnamefont
			{Kalantar-zadeh}},\ }\href {\doibase 10.1016/j.snb.2009.11.007} {\bibfield
		{journal} {\bibinfo  {journal} {Sensors Actuators B Chem.}\ }\textbf
		{\bibinfo {volume} {145}},\ \bibinfo {pages} {13} (\bibinfo {year}
		{2010})}\BibitemShut {NoStop}%
	\bibitem [{\citenamefont {Okumu}\ \emph {et~al.}(2006)\citenamefont {Okumu},
		\citenamefont {Koerfer}, \citenamefont {Salinga}, \citenamefont {Pedersen},\
		and\ \citenamefont {Wuttig}}]{Okumu:2006_ThinSolidFilms}%
	\BibitemOpen
	\bibfield  {author} {\bibinfo {author} {\bibfnamefont {J.}~\bibnamefont
			{Okumu}}, \bibinfo {author} {\bibfnamefont {F.}~\bibnamefont {Koerfer}},
		\bibinfo {author} {\bibfnamefont {C.}~\bibnamefont {Salinga}}, \bibinfo
		{author} {\bibfnamefont {T.~P.}\ \bibnamefont {Pedersen}}, \ and\ \bibinfo
		{author} {\bibfnamefont {M.}~\bibnamefont {Wuttig}},\ }\href {\doibase
		10.1016/j.tsf.2006.03.045} {\bibfield  {journal} {\bibinfo  {journal} {Thin
				Solid Films}\ }\textbf {\bibinfo {volume} {515}},\ \bibinfo {pages} {1327}
		(\bibinfo {year} {2006})}\BibitemShut {NoStop}%
	\bibitem [{\citenamefont {Al-Kuhaili}\ \emph {et~al.}(2015)\citenamefont
		{Al-Kuhaili}, \citenamefont {Ahmad}, \citenamefont {Durrani}, \citenamefont
		{Faiz},\ and\ \citenamefont {Ul-Hamid}}]{AlKuhaili:2015_Mater.Design.}%
	\BibitemOpen
	\bibfield  {author} {\bibinfo {author} {\bibfnamefont {M.~F.}\ \bibnamefont
			{Al-Kuhaili}}, \bibinfo {author} {\bibfnamefont {S.~H.~A.}\ \bibnamefont
			{Ahmad}}, \bibinfo {author} {\bibfnamefont {S.~M.~A.}\ \bibnamefont
			{Durrani}}, \bibinfo {author} {\bibfnamefont {M.~M.}\ \bibnamefont {Faiz}}, \
		and\ \bibinfo {author} {\bibfnamefont {A.}~\bibnamefont {Ul-Hamid}},\ }\href
	{\doibase 10.1016/j.matdes.2015.02.025} {\bibfield  {journal} {\bibinfo
			{journal} {Mater. Des.}\ }\textbf {\bibinfo {volume} {73}},\ \bibinfo {pages}
		{15} (\bibinfo {year} {2015})}\BibitemShut {NoStop}%
	\bibitem [{\citenamefont {Ramana}\ \emph {et~al.}(2012)\citenamefont {Ramana},
		\citenamefont {Atuchin}, \citenamefont {Groult},\ and\ \citenamefont
		{Julien}}]{Ramana:2012_J.Vac.Sci.Technol.A}%
	\BibitemOpen
	\bibfield  {author} {\bibinfo {author} {\bibfnamefont {C.~V.}\ \bibnamefont
			{Ramana}}, \bibinfo {author} {\bibfnamefont {V.~V.}\ \bibnamefont {Atuchin}},
		\bibinfo {author} {\bibfnamefont {H.}~\bibnamefont {Groult}}, \ and\ \bibinfo
		{author} {\bibfnamefont {C.~M.}\ \bibnamefont {Julien}},\ }\href {\doibase
		10.1116/1.3701763} {\bibfield  {journal} {\bibinfo  {journal} {J. Vac. Sci.
				Technol. A Vacuum, Surfaces, Film.}\ }\textbf {\bibinfo {volume} {30}},\
		\bibinfo {pages} {04D105} (\bibinfo {year} {2012})}\BibitemShut {NoStop}%
	\bibitem [{\citenamefont {Nirupama}\ \emph {et~al.}(2010)\citenamefont
		{Nirupama}, \citenamefont {Sekhar}, \citenamefont {Subramanyam},\ and\
		\citenamefont {Uthanna}}]{Rao:2013_Res.J.Recent.Sci.}%
	\BibitemOpen
	\bibfield  {author} {\bibinfo {author} {\bibfnamefont {V.}~\bibnamefont
			{Nirupama}}, \bibinfo {author} {\bibfnamefont {M.~C.}\ \bibnamefont
			{Sekhar}}, \bibinfo {author} {\bibfnamefont {T.~K.}\ \bibnamefont
			{Subramanyam}}, \ and\ \bibinfo {author} {\bibfnamefont {S.}~\bibnamefont
			{Uthanna}},\ }\href {\doibase 10.1088/1742-6596/208/1/012101} {\bibfield
		{journal} {\bibinfo  {journal} {J. Phys. Conf. Ser.}\ }\textbf {\bibinfo
			{volume} {208}},\ \bibinfo {pages} {012101} (\bibinfo {year}
		{2010})}\BibitemShut {NoStop}%
	\bibitem [{\citenamefont {Gesheva}\ \emph {et~al.}(2014)\citenamefont
		{Gesheva}, \citenamefont {Ivanova},\ and\ \citenamefont
		{Bodurov}}]{Gesheva:2014_J.Phys.Conf.Ser.}%
	\BibitemOpen
	\bibfield  {author} {\bibinfo {author} {\bibfnamefont {K.}~\bibnamefont
			{Gesheva}}, \bibinfo {author} {\bibfnamefont {T.}~\bibnamefont {Ivanova}}, \
		and\ \bibinfo {author} {\bibfnamefont {G.}~\bibnamefont {Bodurov}},\ }\href
	{\doibase 10.1088/1742-6596/559/1/012002} {\bibfield  {journal} {\bibinfo
			{journal} {J. Phys. Conf. Ser.}\ }\textbf {\bibinfo {volume} {559}},\
		\bibinfo {pages} {012002} (\bibinfo {year} {2014})}\BibitemShut {NoStop}%
	\bibitem [{\citenamefont {Runnerstrom}\ \emph {et~al.}(2014)\citenamefont
		{Runnerstrom}, \citenamefont {Llord{\'{e}}s}, \citenamefont {Lounis},\ and\
		\citenamefont {Milliron}}]{Runnerstrom:2014_Chem.Commun.}%
	\BibitemOpen
	\bibfield  {author} {\bibinfo {author} {\bibfnamefont {E.~L.}\ \bibnamefont
			{Runnerstrom}}, \bibinfo {author} {\bibfnamefont {A.}~\bibnamefont
			{Llord{\'{e}}s}}, \bibinfo {author} {\bibfnamefont {S.~D.}\ \bibnamefont
			{Lounis}}, \ and\ \bibinfo {author} {\bibfnamefont {D.~J.}\ \bibnamefont
			{Milliron}},\ }\href {\doibase 10.1039/c4cc03109a} {\bibfield  {journal}
		{\bibinfo  {journal} {Chem. Commun. (Camb).}\ }\textbf {\bibinfo {volume}
			{50}},\ \bibinfo {pages} {10555} (\bibinfo {year} {2014})}\BibitemShut
	{NoStop}%
	\bibitem [{\citenamefont {Cao}\ \emph {et~al.}(2015)\citenamefont {Cao},
		\citenamefont {Zheng}, \citenamefont {Shi}, \citenamefont {Yang},
		\citenamefont {Fan}, \citenamefont {Luo}, \citenamefont {Rui}, \citenamefont
		{Chen}, \citenamefont {Yan},\ and\ \citenamefont
		{Zhang}}]{Cao:2015_Adv.Mater.}%
	\BibitemOpen
	\bibfield  {author} {\bibinfo {author} {\bibfnamefont {X.}~\bibnamefont
			{Cao}}, \bibinfo {author} {\bibfnamefont {B.}~\bibnamefont {Zheng}}, \bibinfo
		{author} {\bibfnamefont {W.}~\bibnamefont {Shi}}, \bibinfo {author}
		{\bibfnamefont {J.}~\bibnamefont {Yang}}, \bibinfo {author} {\bibfnamefont
			{Z.}~\bibnamefont {Fan}}, \bibinfo {author} {\bibfnamefont {Z.}~\bibnamefont
			{Luo}}, \bibinfo {author} {\bibfnamefont {X.}~\bibnamefont {Rui}}, \bibinfo
		{author} {\bibfnamefont {B.}~\bibnamefont {Chen}}, \bibinfo {author}
		{\bibfnamefont {Q.}~\bibnamefont {Yan}}, \ and\ \bibinfo {author}
		{\bibfnamefont {H.}~\bibnamefont {Zhang}},\ }\href {\doibase
		10.1002/adma.201501310} {\bibfield  {journal} {\bibinfo  {journal} {Adv.
				Mater.}\ }\textbf {\bibinfo {volume} {27}},\ \bibinfo {pages} {4695}
		(\bibinfo {year} {2015})}\BibitemShut {NoStop}%
	\bibitem [{\citenamefont {Chu}\ \emph {et~al.}(2005)\citenamefont {Chu},
		\citenamefont {Li}, \citenamefont {Chen}, \citenamefont {Shrotriya},\ and\
		\citenamefont {Yang}}]{Chu:2005_Appl.Phys.Lett.}%
	\BibitemOpen
	\bibfield  {author} {\bibinfo {author} {\bibfnamefont {C.~W.}\ \bibnamefont
			{Chu}}, \bibinfo {author} {\bibfnamefont {S.~H.}\ \bibnamefont {Li}},
		\bibinfo {author} {\bibfnamefont {C.~W.}\ \bibnamefont {Chen}}, \bibinfo
		{author} {\bibfnamefont {V.}~\bibnamefont {Shrotriya}}, \ and\ \bibinfo
		{author} {\bibfnamefont {Y.}~\bibnamefont {Yang}},\ }\href {\doibase
		10.1063/1.2126140} {\bibfield  {journal} {\bibinfo  {journal} {Appl. Phys.
				Lett.}\ }\textbf {\bibinfo {volume} {87}},\ \bibinfo {pages} {1} (\bibinfo
		{year} {2005})}\BibitemShut {NoStop}%
	\bibitem [{\citenamefont {Meyer}\ \emph {et~al.}(2012)\citenamefont {Meyer},
		\citenamefont {Hamwi}, \citenamefont {Kr{\"{o}}ger}, \citenamefont
		{Kowalsky}, \citenamefont {Riedl},\ and\ \citenamefont
		{Kahn}}]{Meyer:2012_Adv.Mater.}%
	\BibitemOpen
	\bibfield  {author} {\bibinfo {author} {\bibfnamefont {J.}~\bibnamefont
			{Meyer}}, \bibinfo {author} {\bibfnamefont {S.}~\bibnamefont {Hamwi}},
		\bibinfo {author} {\bibfnamefont {M.}~\bibnamefont {Kr{\"{o}}ger}}, \bibinfo
		{author} {\bibfnamefont {W.}~\bibnamefont {Kowalsky}}, \bibinfo {author}
		{\bibfnamefont {T.}~\bibnamefont {Riedl}}, \ and\ \bibinfo {author}
		{\bibfnamefont {A.}~\bibnamefont {Kahn}},\ }\href {\doibase
		10.1002/adma.201201630} {\bibfield  {journal} {\bibinfo  {journal} {Adv.
				Mater.}\ }\textbf {\bibinfo {volume} {24}},\ \bibinfo {pages} {5408}
		(\bibinfo {year} {2012})}\BibitemShut {NoStop}%
	\bibitem [{\citenamefont {Sun}\ \emph {et~al.}(2011)\citenamefont {Sun},
		\citenamefont {Takacs}, \citenamefont {Cowan}, \citenamefont {Seo},
		\citenamefont {Gong}, \citenamefont {Roy},\ and\ \citenamefont
		{Heeger}}]{Sun:2011_Adv.Mater.}%
	\BibitemOpen
	\bibfield  {author} {\bibinfo {author} {\bibfnamefont {Y.}~\bibnamefont
			{Sun}}, \bibinfo {author} {\bibfnamefont {C.~J.}\ \bibnamefont {Takacs}},
		\bibinfo {author} {\bibfnamefont {S.~R.}\ \bibnamefont {Cowan}}, \bibinfo
		{author} {\bibfnamefont {J.~H.}\ \bibnamefont {Seo}}, \bibinfo {author}
		{\bibfnamefont {X.}~\bibnamefont {Gong}}, \bibinfo {author} {\bibfnamefont
			{A.}~\bibnamefont {Roy}}, \ and\ \bibinfo {author} {\bibfnamefont {A.~J.}\
			\bibnamefont {Heeger}},\ }\href {\doibase 10.1002/adma.201100038} {\bibfield
		{journal} {\bibinfo  {journal} {Adv. Mater.}\ }\textbf {\bibinfo {volume}
			{23}},\ \bibinfo {pages} {2226} (\bibinfo {year} {2011})}\BibitemShut
	{NoStop}%
	\bibitem [{\citenamefont {Battaglia}\ \emph {et~al.}(2014)\citenamefont
		{Battaglia}, \citenamefont {Yin}, \citenamefont {Zheng}, \citenamefont
		{Sharp}, \citenamefont {Chen}, \citenamefont {Mcdonnell}, \citenamefont
		{Azcatl}, \citenamefont {Carraro}, \citenamefont {Ma}, \citenamefont
		{Maboudian}, \citenamefont {Wallace},\ and\ \citenamefont
		{Javey}}]{Battaglia:2014_NanoLett.}%
	\BibitemOpen
	\bibfield  {author} {\bibinfo {author} {\bibfnamefont {C.}~\bibnamefont
			{Battaglia}}, \bibinfo {author} {\bibfnamefont {X.}~\bibnamefont {Yin}},
		\bibinfo {author} {\bibfnamefont {M.}~\bibnamefont {Zheng}}, \bibinfo
		{author} {\bibfnamefont {I.~D.}\ \bibnamefont {Sharp}}, \bibinfo {author}
		{\bibfnamefont {T.}~\bibnamefont {Chen}}, \bibinfo {author} {\bibfnamefont
			{S.}~\bibnamefont {Mcdonnell}}, \bibinfo {author} {\bibfnamefont
			{A.}~\bibnamefont {Azcatl}}, \bibinfo {author} {\bibfnamefont
			{C.}~\bibnamefont {Carraro}}, \bibinfo {author} {\bibfnamefont
			{B.}~\bibnamefont {Ma}}, \bibinfo {author} {\bibfnamefont {R.}~\bibnamefont
			{Maboudian}}, \bibinfo {author} {\bibfnamefont {R.~M.}\ \bibnamefont
			{Wallace}}, \ and\ \bibinfo {author} {\bibfnamefont {A.}~\bibnamefont
			{Javey}},\ }\href {\doibase dx.doi.org/10.1021/nl404389u} {\bibfield
		{journal} {\bibinfo  {journal} {Nano Lett.}\ }\textbf {\bibinfo {volume}
			{14}},\ \bibinfo {pages} {967} (\bibinfo {year} {2014})}\BibitemShut
	{NoStop}%
	\bibitem [{\citenamefont {Dang}\ and\ \citenamefont
		{Singh}(2015)}]{Dang:2015_Sci.Rep.}%
	\BibitemOpen
	\bibfield  {author} {\bibinfo {author} {\bibfnamefont {H.}~\bibnamefont
			{Dang}}\ and\ \bibinfo {author} {\bibfnamefont {V.~P.}\ \bibnamefont
			{Singh}},\ }\href {\doibase 10.1038/srep14859} {\bibfield  {journal}
		{\bibinfo  {journal} {Nat. Publ. Gr.}\ }\textbf {\bibinfo {volume} {5}},\
		\bibinfo {pages} {1} (\bibinfo {year} {2015})}\BibitemShut {NoStop}%
	\bibitem [{\citenamefont {Zhang}\ \emph {et~al.}(2014)\citenamefont {Zhang},
		\citenamefont {Borgschulte}, \citenamefont {Castro}, \citenamefont
		{Crockett}, \citenamefont {Gerecke}, \citenamefont {Deniz}, \citenamefont
		{Heier}, \citenamefont {Jenatsch}, \citenamefont {N{\"{u}}esch},
		\citenamefont {Sanchez-Sanchez}, \citenamefont {Zoladek-Lemanczyk},\ and\
		\citenamefont {Hany}}]{Zhang_2014_Adv.Mater.}%
	\BibitemOpen
	\bibfield  {author} {\bibinfo {author} {\bibfnamefont {H.}~\bibnamefont
			{Zhang}}, \bibinfo {author} {\bibfnamefont {A.}~\bibnamefont {Borgschulte}},
		\bibinfo {author} {\bibfnamefont {F.~A.}\ \bibnamefont {Castro}}, \bibinfo
		{author} {\bibfnamefont {R.}~\bibnamefont {Crockett}}, \bibinfo {author}
		{\bibfnamefont {A.~C.}\ \bibnamefont {Gerecke}}, \bibinfo {author}
		{\bibfnamefont {O.}~\bibnamefont {Deniz}}, \bibinfo {author} {\bibfnamefont
			{J.}~\bibnamefont {Heier}}, \bibinfo {author} {\bibfnamefont
			{S.}~\bibnamefont {Jenatsch}}, \bibinfo {author} {\bibfnamefont
			{F.}~\bibnamefont {N{\"{u}}esch}}, \bibinfo {author} {\bibfnamefont
			{C.}~\bibnamefont {Sanchez-Sanchez}}, \bibinfo {author} {\bibfnamefont
			{A.}~\bibnamefont {Zoladek-Lemanczyk}}, \ and\ \bibinfo {author}
		{\bibfnamefont {R.}~\bibnamefont {Hany}},\ }\href {\doibase
		10.1002/aenm.201400734} {\bibfield  {journal} {\bibinfo  {journal} {Adv.
				Energy Mater.}\ }\textbf {\bibinfo {volume} {5}},\ \bibinfo {pages} {1}
		(\bibinfo {year} {2014})}\BibitemShut {NoStop}%
	\bibitem [{\citenamefont {Zheng}\ \emph {et~al.}(2015)\citenamefont {Zheng},
		\citenamefont {Zhang}, \citenamefont {Zhang}, \citenamefont {Zhao},
		\citenamefont {Ye}, \citenamefont {Chen}, \citenamefont {Yang},\ and\
		\citenamefont {Hou}}]{Zheng:2015_Adv.Mater.}%
	\BibitemOpen
	\bibfield  {author} {\bibinfo {author} {\bibfnamefont {Z.}~\bibnamefont
			{Zheng}}, \bibinfo {author} {\bibfnamefont {S.}~\bibnamefont {Zhang}},
		\bibinfo {author} {\bibfnamefont {M.}~\bibnamefont {Zhang}}, \bibinfo
		{author} {\bibfnamefont {K.}~\bibnamefont {Zhao}}, \bibinfo {author}
		{\bibfnamefont {L.}~\bibnamefont {Ye}}, \bibinfo {author} {\bibfnamefont
			{Y.}~\bibnamefont {Chen}}, \bibinfo {author} {\bibfnamefont {B.}~\bibnamefont
			{Yang}}, \ and\ \bibinfo {author} {\bibfnamefont {J.}~\bibnamefont {Hou}},\
	}\href {\doibase 10.1002/adma.201404525} {\bibfield  {journal} {\bibinfo
		{journal} {Adv. Mater.}\ }\textbf {\bibinfo {volume} {27}},\ \bibinfo {pages}
	{1189} (\bibinfo {year} {2015})}\BibitemShut {NoStop}%
\bibitem [{\citenamefont {Elumalai}\ \emph {et~al.}(2015)\citenamefont
	{Elumalai}, \citenamefont {Vijila}, \citenamefont {Jose}, \citenamefont
	{Uddin},\ and\ \citenamefont
	{Ramakrishna}}]{Elumalai_2015_Mater.Renew.Sustain.Energy}%
\BibitemOpen
\bibfield  {author} {\bibinfo {author} {\bibfnamefont {N.~K.}\ \bibnamefont
		{Elumalai}}, \bibinfo {author} {\bibfnamefont {C.}~\bibnamefont {Vijila}},
	\bibinfo {author} {\bibfnamefont {R.}~\bibnamefont {Jose}}, \bibinfo {author}
	{\bibfnamefont {A.}~\bibnamefont {Uddin}}, \ and\ \bibinfo {author}
	{\bibfnamefont {S.}~\bibnamefont {Ramakrishna}},\ }\href {\doibase
	10.1007/s40243-015-0054-9} {\bibfield  {journal} {\bibinfo  {journal} {Mater.
			Renew. Sustain. Energy}\ }\textbf {\bibinfo {volume} {4}},\ \bibinfo {pages}
	{1} (\bibinfo {year} {2015})}\BibitemShut {NoStop}%
\bibitem [{\citenamefont {Vasilopoulou}\ \emph
	{et~al.}(2012{\natexlab{a}})\citenamefont {Vasilopoulou}, \citenamefont
	{Palilis}, \citenamefont {Georgiadou}, \citenamefont {Kennou}, \citenamefont
	{Kostis}, \citenamefont {Davazoglou},\ and\ \citenamefont
	{Argitis}}]{Vasilopoulou:2012_Appl.Phys.Lett.}%
\BibitemOpen
\bibfield  {author} {\bibinfo {author} {\bibfnamefont {M.}~\bibnamefont
		{Vasilopoulou}}, \bibinfo {author} {\bibfnamefont {L.~C.}\ \bibnamefont
		{Palilis}}, \bibinfo {author} {\bibfnamefont {D.~G.}\ \bibnamefont
		{Georgiadou}}, \bibinfo {author} {\bibfnamefont {S.}~\bibnamefont {Kennou}},
	\bibinfo {author} {\bibfnamefont {I.}~\bibnamefont {Kostis}}, \bibinfo
	{author} {\bibfnamefont {D.}~\bibnamefont {Davazoglou}}, \ and\ \bibinfo
	{author} {\bibfnamefont {P.}~\bibnamefont {Argitis}},\ }\href {\doibase
	10.1063/1.3673283} {\bibfield  {journal} {\bibinfo  {journal} {Appl. Phys.
			Lett.}\ }\textbf {\bibinfo {volume} {100}} (\bibinfo {year}
	{2012}{\natexlab{a}}),\ 10.1063/1.3673283}\BibitemShut {NoStop}%
\bibitem [{\citenamefont {Meyer}\ \emph {et~al.}(2011)\citenamefont {Meyer},
	\citenamefont {Khalandovsky}, \citenamefont {G{\"{o}}rrn},\ and\
	\citenamefont {Kahn}}]{Meyer:2011_Adv.Mat.}%
\BibitemOpen
\bibfield  {author} {\bibinfo {author} {\bibfnamefont {J.}~\bibnamefont
		{Meyer}}, \bibinfo {author} {\bibfnamefont {R.}~\bibnamefont {Khalandovsky}},
	\bibinfo {author} {\bibfnamefont {P.}~\bibnamefont {G{\"{o}}rrn}}, \ and\
	\bibinfo {author} {\bibfnamefont {A.}~\bibnamefont {Kahn}},\ }\href {\doibase
	10.1002/adma.201003065} {\bibfield  {journal} {\bibinfo  {journal} {Adv.
			Mater.}\ }\textbf {\bibinfo {volume} {23}},\ \bibinfo {pages} {70} (\bibinfo
	{year} {2011})}\BibitemShut {NoStop}%
\bibitem [{\citenamefont {Meyer}\ \emph {et~al.}(2014)\citenamefont {Meyer},
	\citenamefont {Kidambi}, \citenamefont {Bayer}, \citenamefont {Weijtens},
	\citenamefont {Kuhn}, \citenamefont {Centeno}, \citenamefont {Pesquera},
	\citenamefont {Zurutuza}, \citenamefont {Robertson},\ and\ \citenamefont
	{Hofmann}}]{Meyer:2014_Sci.Rep.}%
\BibitemOpen
\bibfield  {author} {\bibinfo {author} {\bibfnamefont {J.}~\bibnamefont
		{Meyer}}, \bibinfo {author} {\bibfnamefont {P.~R.}\ \bibnamefont {Kidambi}},
	\bibinfo {author} {\bibfnamefont {B.~C.}\ \bibnamefont {Bayer}}, \bibinfo
	{author} {\bibfnamefont {C.}~\bibnamefont {Weijtens}}, \bibinfo {author}
	{\bibfnamefont {A.}~\bibnamefont {Kuhn}}, \bibinfo {author} {\bibfnamefont
		{A.}~\bibnamefont {Centeno}}, \bibinfo {author} {\bibfnamefont
		{A.}~\bibnamefont {Pesquera}}, \bibinfo {author} {\bibfnamefont
		{A.}~\bibnamefont {Zurutuza}}, \bibinfo {author} {\bibfnamefont
		{J.}~\bibnamefont {Robertson}}, \ and\ \bibinfo {author} {\bibfnamefont
		{S.}~\bibnamefont {Hofmann}},\ }\href {\doibase 10.1038/srep05380} {\bibfield
	{journal} {\bibinfo  {journal} {Sci. Rep.}\ }\textbf {\bibinfo {volume}
		{4}},\ \bibinfo {pages} {5380} (\bibinfo {year} {2014})}\BibitemShut
{NoStop}%
\bibitem [{\citenamefont {Minari}\ \emph {et~al.}(2009)\citenamefont {Minari},
	\citenamefont {Kano}, \citenamefont {Miyadera}, \citenamefont {Wang},
	\citenamefont {Aoyagi},\ and\ \citenamefont
	{Tsukagoshi}}]{Minari:2009_Appl.Phys.Lett.}%
\BibitemOpen
\bibfield  {author} {\bibinfo {author} {\bibfnamefont {T.}~\bibnamefont
		{Minari}}, \bibinfo {author} {\bibfnamefont {M.}~\bibnamefont {Kano}},
	\bibinfo {author} {\bibfnamefont {T.}~\bibnamefont {Miyadera}}, \bibinfo
	{author} {\bibfnamefont {S.~D.}\ \bibnamefont {Wang}}, \bibinfo {author}
	{\bibfnamefont {Y.}~\bibnamefont {Aoyagi}}, \ and\ \bibinfo {author}
	{\bibfnamefont {K.}~\bibnamefont {Tsukagoshi}},\ }\href {\doibase
	10.1063/1.3095665} {\bibfield  {journal} {\bibinfo  {journal} {Appl. Phys.
			Lett.}\ }\textbf {\bibinfo {volume} {94}},\ \bibinfo {pages} {13} (\bibinfo
	{year} {2009})}\BibitemShut {NoStop}%
\bibitem [{\citenamefont {Balendhran}\ \emph
	{et~al.}(2013{\natexlab{a}})\citenamefont {Balendhran}, \citenamefont
	{Walia}, \citenamefont {Nili}, \citenamefont {Ou}, \citenamefont {Zhuiykov},
	\citenamefont {Kaner}, \citenamefont {Sriram}, \citenamefont {Bhaskaran},\
	and\ \citenamefont {Kalantar-zadeh}}]{Balendhran:2013_Adv.Funct.Mater.}%
\BibitemOpen
\bibfield  {author} {\bibinfo {author} {\bibfnamefont {S.}~\bibnamefont
		{Balendhran}}, \bibinfo {author} {\bibfnamefont {S.}~\bibnamefont {Walia}},
	\bibinfo {author} {\bibfnamefont {H.}~\bibnamefont {Nili}}, \bibinfo {author}
	{\bibfnamefont {J.~Z.}\ \bibnamefont {Ou}}, \bibinfo {author} {\bibfnamefont
		{S.}~\bibnamefont {Zhuiykov}}, \bibinfo {author} {\bibfnamefont {R.~B.}\
		\bibnamefont {Kaner}}, \bibinfo {author} {\bibfnamefont {S.}~\bibnamefont
		{Sriram}}, \bibinfo {author} {\bibfnamefont {M.}~\bibnamefont {Bhaskaran}}, \
	and\ \bibinfo {author} {\bibfnamefont {K.}~\bibnamefont {Kalantar-zadeh}},\
}\href {\doibase 10.1002/adfm.201300125} {\bibfield  {journal} {\bibinfo
	{journal} {Adv. Funct. Mater.}\ }\textbf {\bibinfo {volume} {23}},\ \bibinfo
{pages} {3952} (\bibinfo {year} {2013}{\natexlab{a}})}\BibitemShut {NoStop}%
\bibitem [{\citenamefont {Balendhran}\ \emph
	{et~al.}(2013{\natexlab{b}})\citenamefont {Balendhran}, \citenamefont {Deng},
	\citenamefont {Ou}, \citenamefont {Walia}, \citenamefont {Scott},
	\citenamefont {Tang}, \citenamefont {Wang}, \citenamefont {Field},
	\citenamefont {Russo}, \citenamefont {Zhuiykov}, \citenamefont {Strano},
	\citenamefont {Medhekar}, \citenamefont {Sriram}, \citenamefont {Bhaskaran},\
	and\ \citenamefont {Kalantar-zadeh}}]{Balendhran:2013_Adv.Mater.}%
\BibitemOpen
\bibfield  {author} {\bibinfo {author} {\bibfnamefont {S.}~\bibnamefont
		{Balendhran}}, \bibinfo {author} {\bibfnamefont {J.}~\bibnamefont {Deng}},
	\bibinfo {author} {\bibfnamefont {J.~Z.}\ \bibnamefont {Ou}}, \bibinfo
	{author} {\bibfnamefont {S.}~\bibnamefont {Walia}}, \bibinfo {author}
	{\bibfnamefont {J.}~\bibnamefont {Scott}}, \bibinfo {author} {\bibfnamefont
		{J.}~\bibnamefont {Tang}}, \bibinfo {author} {\bibfnamefont {K.~L.}\
		\bibnamefont {Wang}}, \bibinfo {author} {\bibfnamefont {M.~R.}\ \bibnamefont
		{Field}}, \bibinfo {author} {\bibfnamefont {S.}~\bibnamefont {Russo}},
	\bibinfo {author} {\bibfnamefont {S.}~\bibnamefont {Zhuiykov}}, \bibinfo
	{author} {\bibfnamefont {M.~S.}\ \bibnamefont {Strano}}, \bibinfo {author}
	{\bibfnamefont {N.}~\bibnamefont {Medhekar}}, \bibinfo {author}
	{\bibfnamefont {S.}~\bibnamefont {Sriram}}, \bibinfo {author} {\bibfnamefont
		{M.}~\bibnamefont {Bhaskaran}}, \ and\ \bibinfo {author} {\bibfnamefont
		{K.}~\bibnamefont {Kalantar-zadeh}},\ }\href {\doibase
	10.1002/adma.201203346} {\bibfield  {journal} {\bibinfo  {journal} {Adv.
			Mater.}\ }\textbf {\bibinfo {volume} {25}},\ \bibinfo {pages} {109} (\bibinfo
	{year} {2013}{\natexlab{b}})}\BibitemShut {NoStop}%
\bibitem [{\citenamefont {McDonnell}\ \emph {et~al.}(2014)\citenamefont
	{McDonnell}, \citenamefont {Azcatl}, \citenamefont {Addou}, \citenamefont
	{Gong}, \citenamefont {Battaglia}, \citenamefont {Chuang}, \citenamefont
	{Cho}, \citenamefont {Javey},\ and\ \citenamefont
	{Wallace}}]{McDonnell:2014_ACSNano}%
\BibitemOpen
\bibfield  {author} {\bibinfo {author} {\bibfnamefont {S.}~\bibnamefont
		{McDonnell}}, \bibinfo {author} {\bibfnamefont {A.}~\bibnamefont {Azcatl}},
	\bibinfo {author} {\bibfnamefont {R.}~\bibnamefont {Addou}}, \bibinfo
	{author} {\bibfnamefont {C.}~\bibnamefont {Gong}}, \bibinfo {author}
	{\bibfnamefont {C.}~\bibnamefont {Battaglia}}, \bibinfo {author}
	{\bibfnamefont {S.}~\bibnamefont {Chuang}}, \bibinfo {author} {\bibfnamefont
		{K.}~\bibnamefont {Cho}}, \bibinfo {author} {\bibfnamefont {A.}~\bibnamefont
		{Javey}}, \ and\ \bibinfo {author} {\bibfnamefont {R.~M.}\ \bibnamefont
		{Wallace}},\ }\href {\doibase 10.1021/nn501728w} {\bibfield  {journal}
	{\bibinfo  {journal} {ACS Nano}\ }\textbf {\bibinfo {volume} {8}},\ \bibinfo
	{pages} {6265} (\bibinfo {year} {2014})}\BibitemShut {NoStop}%
\bibitem [{\citenamefont {Kostis}\ \emph {et~al.}(2013)\citenamefont {Kostis},
	\citenamefont {Vourdas}, \citenamefont {Papadimitropoulos}, \citenamefont
	{Douvas}, \citenamefont {Vasilopoulou}, \citenamefont {Boukos},\ and\
	\citenamefont {Davazoglou}}]{Kostis:2013_J.Phys.Chem.C}%
\BibitemOpen
\bibfield  {author} {\bibinfo {author} {\bibfnamefont {I.}~\bibnamefont
		{Kostis}}, \bibinfo {author} {\bibfnamefont {N.}~\bibnamefont {Vourdas}},
	\bibinfo {author} {\bibfnamefont {G.}~\bibnamefont {Papadimitropoulos}},
	\bibinfo {author} {\bibfnamefont {A.}~\bibnamefont {Douvas}}, \bibinfo
	{author} {\bibfnamefont {M.}~\bibnamefont {Vasilopoulou}}, \bibinfo {author}
	{\bibfnamefont {N.}~\bibnamefont {Boukos}}, \ and\ \bibinfo {author}
	{\bibfnamefont {D.}~\bibnamefont {Davazoglou}},\ }\href {\doibase
	10.1021/jp407354j} {\bibfield  {journal} {\bibinfo  {journal} {J. Phys. Chem.
			C}\ }\textbf {\bibinfo {volume} {117}},\ \bibinfo {pages} {18013} (\bibinfo
	{year} {2013})}\BibitemShut {NoStop}%
\bibitem [{\citenamefont {Jasieniak}\ \emph {et~al.}(2012)\citenamefont
	{Jasieniak}, \citenamefont {Seifter}, \citenamefont {Jo}, \citenamefont
	{Mates},\ and\ \citenamefont {Heeger}}]{Jasieniak:2012_Adv.Funct.Mater.}%
\BibitemOpen
\bibfield  {author} {\bibinfo {author} {\bibfnamefont {J.~J.}\ \bibnamefont
		{Jasieniak}}, \bibinfo {author} {\bibfnamefont {J.}~\bibnamefont {Seifter}},
	\bibinfo {author} {\bibfnamefont {J.}~\bibnamefont {Jo}}, \bibinfo {author}
	{\bibfnamefont {T.}~\bibnamefont {Mates}}, \ and\ \bibinfo {author}
	{\bibfnamefont {A.~J.}\ \bibnamefont {Heeger}},\ }\href {\doibase
	10.1002/adfm.201102622} {\bibfield  {journal} {\bibinfo  {journal} {Adv.
			Funct. Mater.}\ }\textbf {\bibinfo {volume} {22}},\ \bibinfo {pages} {2594}
	(\bibinfo {year} {2012})}\BibitemShut {NoStop}%
\bibitem [{\citenamefont {Murase}\ and\ \citenamefont
	{Yang}(2012)}]{Murase:2012_Adv.Mater.}%
\BibitemOpen
\bibfield  {author} {\bibinfo {author} {\bibfnamefont {S.}~\bibnamefont
		{Murase}}\ and\ \bibinfo {author} {\bibfnamefont {Y.}~\bibnamefont {Yang}},\
}\href {\doibase 10.1002/adma.201104771} {\bibfield  {journal} {\bibinfo
	{journal} {Adv. Mater.}\ }\textbf {\bibinfo {volume} {24}},\ \bibinfo {pages}
{2459} (\bibinfo {year} {2012})}\BibitemShut {NoStop}%
\bibitem [{\citenamefont {Bao}\ \emph {et~al.}(2010)\citenamefont {Bao},
	\citenamefont {Yang}, \citenamefont {Li},\ and\ \citenamefont
	{Tang}}]{Bao:2010_Appl.Phys.Lett.}%
\BibitemOpen
\bibfield  {author} {\bibinfo {author} {\bibfnamefont {Q.~Y.}\ \bibnamefont
		{Bao}}, \bibinfo {author} {\bibfnamefont {J.~P.}\ \bibnamefont {Yang}},
	\bibinfo {author} {\bibfnamefont {Y.~Q.}\ \bibnamefont {Li}}, \ and\ \bibinfo
	{author} {\bibfnamefont {J.~X.}\ \bibnamefont {Tang}},\ }\href {\doibase
	10.1063/1.3479477} {\bibfield  {journal} {\bibinfo  {journal} {Appl. Phys.
			Lett.}\ }\textbf {\bibinfo {volume} {97}},\ \bibinfo {pages} {063303}
	(\bibinfo {year} {2010})}\BibitemShut {NoStop}%
\bibitem [{\citenamefont {Greiner}\ \emph {et~al.}(2011)\citenamefont
	{Greiner}, \citenamefont {Helander}, \citenamefont {Tang}, \citenamefont
	{Wang}, \citenamefont {Qiu},\ and\ \citenamefont
	{Lu}}]{Greiner:2011_Nat.Mater.}%
\BibitemOpen
\bibfield  {author} {\bibinfo {author} {\bibfnamefont {M.~T.}\ \bibnamefont
		{Greiner}}, \bibinfo {author} {\bibfnamefont {M.~G.}\ \bibnamefont
		{Helander}}, \bibinfo {author} {\bibfnamefont {W.~M.}\ \bibnamefont {Tang}},
	\bibinfo {author} {\bibfnamefont {Z.~B.}\ \bibnamefont {Wang}}, \bibinfo
	{author} {\bibfnamefont {J.}~\bibnamefont {Qiu}}, \ and\ \bibinfo {author}
	{\bibfnamefont {Z.~H.}\ \bibnamefont {Lu}},\ }\href {\doibase
	10.1038/nmat3159} {\bibfield  {journal} {\bibinfo  {journal} {Nat. Mater.}\
	}\textbf {\bibinfo {volume} {11}},\ \bibinfo {pages} {76} (\bibinfo {year}
	{2011})}\BibitemShut {NoStop}%
\bibitem [{\citenamefont {Wong}\ \emph {et~al.}(2012)\citenamefont {Wong},
	\citenamefont {Krishnamoorthy}, \citenamefont {Luther},\ and\ \citenamefont
	{Balaya}}]{Wong:2012_J.Phys.Chem.C}%
\BibitemOpen
\bibfield  {author} {\bibinfo {author} {\bibfnamefont {K.~H.}\ \bibnamefont
		{Wong}}, \bibinfo {author} {\bibfnamefont {A.}~\bibnamefont
		{Krishnamoorthy}}, \bibinfo {author} {\bibfnamefont {J.}~\bibnamefont
		{Luther}}, \ and\ \bibinfo {author} {\bibfnamefont {P.}~\bibnamefont
		{Balaya}},\ }\href {\doibase 10.1021/jp303679y} {\bibfield  {journal}
	{\bibinfo  {journal} {J. Phys. Chem. C}\ }\textbf {\bibinfo {volume} {116}},\
	\bibinfo {pages} {16346} (\bibinfo {year} {2012})}\BibitemShut {NoStop}%
\bibitem [{\citenamefont {Greiner}\ \emph {et~al.}(2012)\citenamefont
	{Greiner}, \citenamefont {Chai}, \citenamefont {Helander}, \citenamefont
	{Tang},\ and\ \citenamefont {Lu}}]{Greiner:2012_Adv.Funct.Mater.}%
\BibitemOpen
\bibfield  {author} {\bibinfo {author} {\bibfnamefont {M.~T.}\ \bibnamefont
		{Greiner}}, \bibinfo {author} {\bibfnamefont {L.}~\bibnamefont {Chai}},
	\bibinfo {author} {\bibfnamefont {M.~G.}\ \bibnamefont {Helander}}, \bibinfo
	{author} {\bibfnamefont {W.~M.}\ \bibnamefont {Tang}}, \ and\ \bibinfo
	{author} {\bibfnamefont {Z.~H.}\ \bibnamefont {Lu}},\ }\href {\doibase
	10.1002/adfm.201200615} {\bibfield  {journal} {\bibinfo  {journal} {Adv.
			Funct. Mater.}\ }\textbf {\bibinfo {volume} {22}},\ \bibinfo {pages} {4557}
	(\bibinfo {year} {2012})}\BibitemShut {NoStop}%
\bibitem [{\citenamefont
	{Magn{\'{e}}li}(1948)}]{Magneli:1948_Acta.Chem.Scand.}%
\BibitemOpen
\bibfield  {author} {\bibinfo {author} {\bibfnamefont {A.}~\bibnamefont
		{Magn{\'{e}}li}},\ }\href {\doibase 10.3891/acta.chem.scand.02-0501}
{\bibfield  {journal} {\bibinfo  {journal} {Acta Chem. Scand.}\ }\textbf
	{\bibinfo {volume} {2}},\ \bibinfo {pages} {501} (\bibinfo {year}
	{1948})}\BibitemShut {NoStop}%
\bibitem [{\citenamefont {Kihlborg}\ \emph {et~al.}(1959)\citenamefont
	{Kihlborg}, \citenamefont {Sundholm}, \citenamefont {Magn{\'{e}}li},
	\citenamefont {H{\"{o}}gberg}, \citenamefont {Kneip},\ and\ \citenamefont
	{Palmstierna}}]{Kihlborg:1959_Acta.Chem.Scand.}%
\BibitemOpen
\bibfield  {author} {\bibinfo {author} {\bibfnamefont {L.}~\bibnamefont
		{Kihlborg}}, \bibinfo {author} {\bibfnamefont {A.}~\bibnamefont {Sundholm}},
	\bibinfo {author} {\bibfnamefont {A.}~\bibnamefont {Magn{\'{e}}li}}, \bibinfo
	{author} {\bibfnamefont {B.}~\bibnamefont {H{\"{o}}gberg}}, \bibinfo {author}
	{\bibfnamefont {P.}~\bibnamefont {Kneip}}, \ and\ \bibinfo {author}
	{\bibfnamefont {H.}~\bibnamefont {Palmstierna}},\ }\href {\doibase
	10.3891/acta.chem.scand.13-0954} {\bibfield  {journal} {\bibinfo  {journal}
		{Acta Chem. Scand.}\ }\textbf {\bibinfo {volume} {13}},\ \bibinfo {pages}
	{954} (\bibinfo {year} {1959})}\BibitemShut {NoStop}%
\bibitem [{\citenamefont {Irfan}\ and\ \citenamefont
	{Gao}(2012)}]{Ifran:2012_J.Photon.Energy}%
\BibitemOpen
\bibfield  {author} {\bibinfo {author} {\bibfnamefont {I.}~\bibnamefont
		{Irfan}}\ and\ \bibinfo {author} {\bibfnamefont {Y.}~\bibnamefont {Gao}},\
}\href {\doibase 10.1117/1.JPE.2.021213} {\bibfield  {journal} {\bibinfo
	{journal} {J. Photonics Energy}\ }\textbf {\bibinfo {volume} {2}},\ \bibinfo
{pages} {021213} (\bibinfo {year} {2012})}\BibitemShut {NoStop}%
\bibitem [{\citenamefont {Irfan}\ \emph {et~al.}(2012)\citenamefont {Irfan},
	\citenamefont {{James Turinske}}, \citenamefont {Bao},\ and\ \citenamefont
	{Gao}}]{Ifran:2012_Appl.Phys.Lett.}%
\BibitemOpen
\bibfield  {author} {\bibinfo {author} {\bibfnamefont {I.}~\bibnamefont
		{Irfan}}, \bibinfo {author} {\bibfnamefont {A.}~\bibnamefont {{James
				Turinske}}}, \bibinfo {author} {\bibfnamefont {Z.}~\bibnamefont {Bao}}, \
	and\ \bibinfo {author} {\bibfnamefont {Y.}~\bibnamefont {Gao}},\ }\href
{\doibase 10.1063/1.4748978} {\bibfield  {journal} {\bibinfo  {journal}
		{Appl. Phys. Lett.}\ }\textbf {\bibinfo {volume} {101}} (\bibinfo {year}
	{2012}),\ 10.1063/1.4748978}\BibitemShut {NoStop}%
\bibitem [{\citenamefont {Huang}\ \emph {et~al.}(2014)\citenamefont {Huang},
	\citenamefont {He}, \citenamefont {Cao},\ and\ \citenamefont
	{Lu}}]{Huang:2014_Sci.Rep.}%
\BibitemOpen
\bibfield  {author} {\bibinfo {author} {\bibfnamefont {P.-R.}\ \bibnamefont
		{Huang}}, \bibinfo {author} {\bibfnamefont {Y.}~\bibnamefont {He}}, \bibinfo
	{author} {\bibfnamefont {C.}~\bibnamefont {Cao}}, \ and\ \bibinfo {author}
	{\bibfnamefont {Z.-H.}\ \bibnamefont {Lu}},\ }\href {\doibase
	10.1038/srep07131} {\bibfield  {journal} {\bibinfo  {journal} {Sci. Rep.}\
	}\textbf {\bibinfo {volume} {4}},\ \bibinfo {pages} {7131} (\bibinfo {year}
	{2014})}\BibitemShut {NoStop}%
\bibitem [{\citenamefont {Kr{\"{o}}ger}\ \emph {et~al.}(2009)\citenamefont
	{Kr{\"{o}}ger}, \citenamefont {Hamwi}, \citenamefont {Meyer}, \citenamefont
	{Riedl}, \citenamefont {Kowalsky},\ and\ \citenamefont
	{Kahn}}]{Kroger:2009_Appl.Phys.Lett}%
\BibitemOpen
\bibfield  {author} {\bibinfo {author} {\bibfnamefont {M.}~\bibnamefont
		{Kr{\"{o}}ger}}, \bibinfo {author} {\bibfnamefont {S.}~\bibnamefont {Hamwi}},
	\bibinfo {author} {\bibfnamefont {J.}~\bibnamefont {Meyer}}, \bibinfo
	{author} {\bibfnamefont {T.}~\bibnamefont {Riedl}}, \bibinfo {author}
	{\bibfnamefont {W.}~\bibnamefont {Kowalsky}}, \ and\ \bibinfo {author}
	{\bibfnamefont {A.}~\bibnamefont {Kahn}},\ }\href {\doibase
	10.1063/1.3231928} {\bibfield  {journal} {\bibinfo  {journal} {Appl. Phys.
			Lett.}\ }\textbf {\bibinfo {volume} {95}},\ \bibinfo {pages} {100} (\bibinfo
	{year} {2009})}\BibitemShut {NoStop}%
\bibitem [{\citenamefont {Irfan}\ \emph {et~al.}(2010)\citenamefont {Irfan},
	\citenamefont {Ding}, \citenamefont {Gao}, \citenamefont {Small},
	\citenamefont {Kim}, \citenamefont {Subbiah},\ and\ \citenamefont
	{So}}]{Ifran_2010_Appl.Phys.Lett.}%
\BibitemOpen
\bibfield  {author} {\bibinfo {author} {\bibnamefont {Irfan}}, \bibinfo
	{author} {\bibfnamefont {H.}~\bibnamefont {Ding}}, \bibinfo {author}
	{\bibfnamefont {Y.}~\bibnamefont {Gao}}, \bibinfo {author} {\bibfnamefont
		{C.}~\bibnamefont {Small}}, \bibinfo {author} {\bibfnamefont {D.~Y.}\
		\bibnamefont {Kim}}, \bibinfo {author} {\bibfnamefont {J.}~\bibnamefont
		{Subbiah}}, \ and\ \bibinfo {author} {\bibfnamefont {F.}~\bibnamefont {So}},\
}\href {\doibase 10.1063/1.3454779} {\bibfield  {journal} {\bibinfo
	{journal} {Appl. Phys. Lett.}\ }\textbf {\bibinfo {volume} {96}},\ \bibinfo
{pages} {2014} (\bibinfo {year} {2010})}\BibitemShut {NoStop}%
\bibitem [{\citenamefont {Vasilopoulou}\ \emph
	{et~al.}(2012{\natexlab{b}})\citenamefont {Vasilopoulou}, \citenamefont
	{Douvas}, \citenamefont {Georgiadou}, \citenamefont {Palilis}, \citenamefont
	{Kennou}, \citenamefont {Sygellou}, \citenamefont {Soultati}, \citenamefont
	{Kostis}, \citenamefont {Papadimitropoulos}, \citenamefont {Davazoglou},\
	and\ \citenamefont {Argitis}}]{Vasilopoulou:2012_J.Am.Chem.Soc.}%
\BibitemOpen
\bibfield  {author} {\bibinfo {author} {\bibfnamefont {M.}~\bibnamefont
		{Vasilopoulou}}, \bibinfo {author} {\bibfnamefont {A.~M.}\ \bibnamefont
		{Douvas}}, \bibinfo {author} {\bibfnamefont {D.~G.}\ \bibnamefont
		{Georgiadou}}, \bibinfo {author} {\bibfnamefont {L.~C.}\ \bibnamefont
		{Palilis}}, \bibinfo {author} {\bibfnamefont {S.}~\bibnamefont {Kennou}},
	\bibinfo {author} {\bibfnamefont {L.}~\bibnamefont {Sygellou}}, \bibinfo
	{author} {\bibfnamefont {A.}~\bibnamefont {Soultati}}, \bibinfo {author}
	{\bibfnamefont {I.}~\bibnamefont {Kostis}}, \bibinfo {author} {\bibfnamefont
		{G.}~\bibnamefont {Papadimitropoulos}}, \bibinfo {author} {\bibfnamefont
		{D.}~\bibnamefont {Davazoglou}}, \ and\ \bibinfo {author} {\bibfnamefont
		{P.}~\bibnamefont {Argitis}},\ }\href {\doibase 10.1021/ja3026906} {\bibfield
	{journal} {\bibinfo  {journal} {J. Am. Chem. Soc.}\ }\textbf {\bibinfo
		{volume} {134}},\ \bibinfo {pages} {16178} (\bibinfo {year}
	{2012}{\natexlab{b}})}\BibitemShut {NoStop}%
\bibitem [{\citenamefont {Vasilopoulou}\ \emph {et~al.}(2014)\citenamefont
	{Vasilopoulou}, \citenamefont {Soultati}, \citenamefont {Argitis},\ and\
	\citenamefont {Stergiopoulos}}]{Vasilopoulou_2014_J.Phys.Chem.Lett.}%
\BibitemOpen
\bibfield  {author} {\bibinfo {author} {\bibfnamefont {M.}~\bibnamefont
		{Vasilopoulou}}, \bibinfo {author} {\bibfnamefont {A.}~\bibnamefont
		{Soultati}}, \bibinfo {author} {\bibfnamefont {P.}~\bibnamefont {Argitis}}, \
	and\ \bibinfo {author} {\bibfnamefont {T.}~\bibnamefont {Stergiopoulos}},\
}\href {\doibase dx.doi.org/10.1021/jz500612p} {\bibfield  {journal}
{\bibinfo  {journal} {J. Phys. Chem. Lett.}\ }\textbf {\bibinfo {volume}
	{5}},\ \bibinfo {pages} {1871} (\bibinfo {year} {2014})}\BibitemShut
{NoStop}%
\bibitem [{\citenamefont {Soultati}\ \emph {et~al.}(2016)\citenamefont
	{Soultati}, \citenamefont {Kostis}, \citenamefont {Argitis}, \citenamefont
	{Dimotikali}, \citenamefont {Kennou}, \citenamefont {Gardelis}, \citenamefont
	{Speliotis}, \citenamefont {Kontos}, \citenamefont {Davazoglou},\ and\
	\citenamefont {Vasilopoulou}}]{Soultati:2016_J.Phys.Chem.C}%
\BibitemOpen
\bibfield  {author} {\bibinfo {author} {\bibfnamefont {A.}~\bibnamefont
		{Soultati}}, \bibinfo {author} {\bibfnamefont {I.}~\bibnamefont {Kostis}},
	\bibinfo {author} {\bibfnamefont {P.}~\bibnamefont {Argitis}}, \bibinfo
	{author} {\bibfnamefont {D.}~\bibnamefont {Dimotikali}}, \bibinfo {author}
	{\bibfnamefont {S.}~\bibnamefont {Kennou}}, \bibinfo {author} {\bibfnamefont
		{S.}~\bibnamefont {Gardelis}}, \bibinfo {author} {\bibfnamefont
		{T.}~\bibnamefont {Speliotis}}, \bibinfo {author} {\bibfnamefont {A.~G.}\
		\bibnamefont {Kontos}}, \bibinfo {author} {\bibfnamefont {D.}~\bibnamefont
		{Davazoglou}}, \ and\ \bibinfo {author} {\bibfnamefont {M.}~\bibnamefont
		{Vasilopoulou}},\ }\href {\doibase 10.1039/C6TC02259F} {\bibfield  {journal}
	{\bibinfo  {journal} {J. Mater. Chem. C}\ ,\ \bibinfo {pages} {8}} (\bibinfo
	{year} {2016})}\BibitemShut {NoStop}%
\bibitem [{\citenamefont {Bovill}\ \emph {et~al.}(2013)\citenamefont {Bovill},
	\citenamefont {Griffin}, \citenamefont {Wang}, \citenamefont {Kingsley},
	\citenamefont {Yi}, \citenamefont {Iraqi}, \citenamefont {Buckley},\ and\
	\citenamefont {Lidzey}}]{Bovill:2013_Appl.Phys.Lett.}%
\BibitemOpen
\bibfield  {author} {\bibinfo {author} {\bibfnamefont {E.~S.~R.}\
		\bibnamefont {Bovill}}, \bibinfo {author} {\bibfnamefont {J.}~\bibnamefont
		{Griffin}}, \bibinfo {author} {\bibfnamefont {T.}~\bibnamefont {Wang}},
	\bibinfo {author} {\bibfnamefont {J.~W.}\ \bibnamefont {Kingsley}}, \bibinfo
	{author} {\bibfnamefont {H.}~\bibnamefont {Yi}}, \bibinfo {author}
	{\bibfnamefont {A.}~\bibnamefont {Iraqi}}, \bibinfo {author} {\bibfnamefont
		{A.~R.}\ \bibnamefont {Buckley}}, \ and\ \bibinfo {author} {\bibfnamefont
		{D.~G.}\ \bibnamefont {Lidzey}},\ }\href {\doibase 10.1063/1.4804294}
{\bibfield  {journal} {\bibinfo  {journal} {Appl. Phys. Lett.}\ }\textbf
	{\bibinfo {volume} {102}},\ \bibinfo {pages} {183303} (\bibinfo {year}
	{2013})}\BibitemShut {NoStop}%
\bibitem [{\citenamefont {Meyer}\ \emph
	{et~al.}(2010{\natexlab{a}})\citenamefont {Meyer}, \citenamefont {Shu},
	\citenamefont {Kröger},\ and\ \citenamefont
	{Kahn}}]{Meyer:2010b_Appl.Phys.Lett.}%
\BibitemOpen
\bibfield  {author} {\bibinfo {author} {\bibfnamefont {J.}~\bibnamefont
		{Meyer}}, \bibinfo {author} {\bibfnamefont {A.}~\bibnamefont {Shu}}, \bibinfo
	{author} {\bibfnamefont {M.}~\bibnamefont {Kröger}}, \ and\ \bibinfo
	{author} {\bibfnamefont {A.}~\bibnamefont {Kahn}},\ }\href {\doibase
	10.1063/1.3374333} {\bibfield  {journal} {\bibinfo  {journal} {Appl. Phys.
			Lett.}\ }\textbf {\bibinfo {volume} {96}},\ \bibinfo {pages} {133308}
	(\bibinfo {year} {2010}{\natexlab{a}})}\BibitemShut {NoStop}%
\bibitem [{\citenamefont {Lee}\ \emph {et~al.}(2014)\citenamefont {Lee},
	\citenamefont {Liu},\ and\ \citenamefont {Kelly}}]{Lee:2014_J.Phys.Chem.}%
\BibitemOpen
\bibfield  {author} {\bibinfo {author} {\bibfnamefont {K.~E.}\ \bibnamefont
		{Lee}}, \bibinfo {author} {\bibfnamefont {L.}~\bibnamefont {Liu}}, \ and\
	\bibinfo {author} {\bibfnamefont {T.~L.}\ \bibnamefont {Kelly}},\ }\href
{\doibase 10.1021/jp508972v} {\bibfield  {journal} {\bibinfo  {journal} {J.
			Phys. Chem. C}\ }\textbf {\bibinfo {volume} {118}},\ \bibinfo {pages} {27735}
	(\bibinfo {year} {2014})}\BibitemShut {NoStop}%
\bibitem [{\citenamefont {Butler}\ \emph {et~al.}(2015)\citenamefont {Butler},
	\citenamefont {Crespo-Otero}, \citenamefont {Buckeridge}, \citenamefont
	{Scanlon}, \citenamefont {Bovill}, \citenamefont {Lidzey},\ and\
	\citenamefont {Walsh}}]{Butler:2015_Appl.Phys.Lett.}%
\BibitemOpen
\bibfield  {author} {\bibinfo {author} {\bibfnamefont {K.~T.}\ \bibnamefont
		{Butler}}, \bibinfo {author} {\bibfnamefont {R.}~\bibnamefont
		{Crespo-Otero}}, \bibinfo {author} {\bibfnamefont {J.}~\bibnamefont
		{Buckeridge}}, \bibinfo {author} {\bibfnamefont {D.~O.}\ \bibnamefont
		{Scanlon}}, \bibinfo {author} {\bibfnamefont {E.}~\bibnamefont {Bovill}},
	\bibinfo {author} {\bibfnamefont {D.}~\bibnamefont {Lidzey}}, \ and\ \bibinfo
	{author} {\bibfnamefont {A.}~\bibnamefont {Walsh}},\ }\href {\doibase
	10.1063/1.4937460} {\bibfield  {journal} {\bibinfo  {journal} {Appl. Phys.
			Lett.}\ }\textbf {\bibinfo {volume} {107}},\ \bibinfo {pages} {231605}
	(\bibinfo {year} {2015})}\BibitemShut {NoStop}%
\bibitem [{\citenamefont {Kanai}\ \emph {et~al.}(2010)\citenamefont {Kanai},
	\citenamefont {Koizumi}, \citenamefont {Ouchi}, \citenamefont {Tsukamoto},
	\citenamefont {Sakanoue}, \citenamefont {Ouchi},\ and\ \citenamefont
	{Seki}}]{Kanai:2010_Org.Electron.}%
\BibitemOpen
\bibfield  {author} {\bibinfo {author} {\bibfnamefont {K.}~\bibnamefont
		{Kanai}}, \bibinfo {author} {\bibfnamefont {K.}~\bibnamefont {Koizumi}},
	\bibinfo {author} {\bibfnamefont {S.}~\bibnamefont {Ouchi}}, \bibinfo
	{author} {\bibfnamefont {Y.}~\bibnamefont {Tsukamoto}}, \bibinfo {author}
	{\bibfnamefont {K.}~\bibnamefont {Sakanoue}}, \bibinfo {author}
	{\bibfnamefont {Y.}~\bibnamefont {Ouchi}}, \ and\ \bibinfo {author}
	{\bibfnamefont {K.}~\bibnamefont {Seki}},\ }\href {\doibase
	10.1016/j.orgel.2009.10.013} {\bibfield  {journal} {\bibinfo  {journal} {Org.
			Electron. physics, Mater. Appl.}\ }\textbf {\bibinfo {volume} {11}},\
	\bibinfo {pages} {188} (\bibinfo {year} {2010})}\BibitemShut {NoStop}%
\bibitem [{\citenamefont {Meyer}\ \emph
	{et~al.}(2010{\natexlab{b}})\citenamefont {Meyer}, \citenamefont {Kröger},
	\citenamefont {Hamwi}, \citenamefont {Gnam}, \citenamefont {Riedl},
	\citenamefont {Kowalsky},\ and\ \citenamefont
	{Kahn}}]{Meyer:2010_Appl.Phys.Lett.}%
\BibitemOpen
\bibfield  {author} {\bibinfo {author} {\bibfnamefont {J.}~\bibnamefont
		{Meyer}}, \bibinfo {author} {\bibfnamefont {M.}~\bibnamefont {Kröger}},
	\bibinfo {author} {\bibfnamefont {S.}~\bibnamefont {Hamwi}}, \bibinfo
	{author} {\bibfnamefont {F.}~\bibnamefont {Gnam}}, \bibinfo {author}
	{\bibfnamefont {T.}~\bibnamefont {Riedl}}, \bibinfo {author} {\bibfnamefont
		{W.}~\bibnamefont {Kowalsky}}, \ and\ \bibinfo {author} {\bibfnamefont
		{A.}~\bibnamefont {Kahn}},\ }\href {\doibase 10.1063/1.3427430} {\bibfield
	{journal} {\bibinfo  {journal} {Appl. Phys. Lett.}\ }\textbf {\bibinfo
		{volume} {96}},\ \bibinfo {pages} {193302} (\bibinfo {year}
	{2010}{\natexlab{b}})}\BibitemShut {NoStop}%
\bibitem [{\citenamefont {Bj{\"{o}}rck}\ and\ \citenamefont
	{Andersson}(2007)}]{Bjorck:2007_J.Appl.Crystallogr.}%
\BibitemOpen
\bibfield  {author} {\bibinfo {author} {\bibfnamefont {M.}~\bibnamefont
		{Bj{\"{o}}rck}}\ and\ \bibinfo {author} {\bibfnamefont {G.}~\bibnamefont
		{Andersson}},\ }\href {\doibase 10.1107/S0021889807045086} {\bibfield
	{journal} {\bibinfo  {journal} {J. Appl. Crystallogr.}\ }\textbf {\bibinfo
		{volume} {40}},\ \bibinfo {pages} {1174} (\bibinfo {year}
	{2007})}\BibitemShut {NoStop}%
\bibitem [{\citenamefont {Baltrusaitis}\ \emph {et~al.}(2015)\citenamefont
	{Baltrusaitis}, \citenamefont {Mendoza-Sanchez}, \citenamefont {Fernandez},
	\citenamefont {Veenstra}, \citenamefont {Dukstiene}, \citenamefont
	{Roberts},\ and\ \citenamefont {Fairley}}]{Baltrusaitis:2015_Appl.Surf.Sci.}%
\BibitemOpen
\bibfield  {author} {\bibinfo {author} {\bibfnamefont {J.}~\bibnamefont
		{Baltrusaitis}}, \bibinfo {author} {\bibfnamefont {B.}~\bibnamefont
		{Mendoza-Sanchez}}, \bibinfo {author} {\bibfnamefont {V.}~\bibnamefont
		{Fernandez}}, \bibinfo {author} {\bibfnamefont {R.}~\bibnamefont {Veenstra}},
	\bibinfo {author} {\bibfnamefont {N.}~\bibnamefont {Dukstiene}}, \bibinfo
	{author} {\bibfnamefont {A.}~\bibnamefont {Roberts}}, \ and\ \bibinfo
	{author} {\bibfnamefont {N.}~\bibnamefont {Fairley}},\ }\href {\doibase
	10.1016/j.apsusc.2014.11.077} {\bibfield  {journal} {\bibinfo  {journal}
		{Appl. Surf. Sci.}\ }\textbf {\bibinfo {volume} {326}},\ \bibinfo {pages}
	{151} (\bibinfo {year} {2015})}\BibitemShut {NoStop}%
\bibitem [{\citenamefont {Werfel}\ and\ \citenamefont
	{Minni}(1983)}]{Werfel:1983_J.Phys.C.SolidStatePhys.}%
\BibitemOpen
\bibfield  {author} {\bibinfo {author} {\bibfnamefont {F.}~\bibnamefont
		{Werfel}}\ and\ \bibinfo {author} {\bibfnamefont {E.}~\bibnamefont {Minni}},\
}\href {\doibase http://dx.doi.org/10.1088/0022-3719/16/31/022} {\bibfield
{journal} {\bibinfo  {journal} {J. Phys. C Solid State Phys.}\ }\textbf
{\bibinfo {volume} {16}},\ \bibinfo {pages} {6091} (\bibinfo {year}
{1983})}\BibitemShut {NoStop}%
\bibitem [{\citenamefont {Papadopoulos}\ \emph {et~al.}(2013)\citenamefont
	{Papadopoulos}, \citenamefont {Meyer}, \citenamefont {Li}, \citenamefont
	{Guan}, \citenamefont {Kahn},\ and\ \citenamefont
	{Br{\'{e}}das}}]{Papadopoulos:2013_Adv.Funct.Mater.}%
\BibitemOpen
\bibfield  {author} {\bibinfo {author} {\bibfnamefont {T.~A.}\ \bibnamefont
		{Papadopoulos}}, \bibinfo {author} {\bibfnamefont {J.}~\bibnamefont {Meyer}},
	\bibinfo {author} {\bibfnamefont {H.}~\bibnamefont {Li}}, \bibinfo {author}
	{\bibfnamefont {Z.}~\bibnamefont {Guan}}, \bibinfo {author} {\bibfnamefont
		{A.}~\bibnamefont {Kahn}}, \ and\ \bibinfo {author} {\bibfnamefont {J.-L.}\
		\bibnamefont {Br{\'{e}}das}},\ }\href {\doibase 10.1002/adfm.201301466}
{\bibfield  {journal} {\bibinfo  {journal} {Adv. Funct. Mater.}\ }\textbf
	{\bibinfo {volume} {23}},\ \bibinfo {pages} {6091} (\bibinfo {year}
	{2013})}\BibitemShut {NoStop}%
\bibitem [{\citenamefont {Gwinner}\ \emph {et~al.}(2011)\citenamefont
	{Gwinner}, \citenamefont {Pietro}, \citenamefont {Vaynzof}, \citenamefont
	{Greenberg}, \citenamefont {Ho}, \citenamefont {Friend},\ and\ \citenamefont
	{Sirringhaus}}]{Gwinner:2011_Adv.Funct.Mater.}%
\BibitemOpen
\bibfield  {author} {\bibinfo {author} {\bibfnamefont {M.~C.}\ \bibnamefont
		{Gwinner}}, \bibinfo {author} {\bibfnamefont {R.~D.}\ \bibnamefont {Pietro}},
	\bibinfo {author} {\bibfnamefont {Y.}~\bibnamefont {Vaynzof}}, \bibinfo
	{author} {\bibfnamefont {K.~J.}\ \bibnamefont {Greenberg}}, \bibinfo {author}
	{\bibfnamefont {P.~K.~H.}\ \bibnamefont {Ho}}, \bibinfo {author}
	{\bibfnamefont {R.~H.}\ \bibnamefont {Friend}}, \ and\ \bibinfo {author}
	{\bibfnamefont {H.}~\bibnamefont {Sirringhaus}},\ }\href {\doibase
	10.1002/adfm.201002696} {\bibfield  {journal} {\bibinfo  {journal} {Adv.
			Funct. Mater.}\ }\textbf {\bibinfo {volume} {21}},\ \bibinfo {pages} {1432}
	(\bibinfo {year} {2011})}\BibitemShut {NoStop}%
\bibitem [{\citenamefont {Rolandi}\ \emph {et~al.}(2002)\citenamefont
	{Rolandi}, \citenamefont {Quate},\ and\ \citenamefont
	{Dai}}]{Rolandi:2002_Adv.Mater.}%
\BibitemOpen
\bibfield  {author} {\bibinfo {author} {\bibfnamefont {M.}~\bibnamefont
		{Rolandi}}, \bibinfo {author} {\bibfnamefont {C.}~\bibnamefont {Quate}}, \
	and\ \bibinfo {author} {\bibfnamefont {H.}~\bibnamefont {Dai}},\ }\href
{\doibase 10.1002/1521-4095(20020205)14:3<191::AID-ADMA191>3.0.CO;2-7}
{\bibfield  {journal} {\bibinfo  {journal} {Adv. Mater.}\ }\textbf {\bibinfo
		{volume} {14}},\ \bibinfo {pages} {191} (\bibinfo {year} {2002})}\BibitemShut
{NoStop}%
\bibitem [{\citenamefont {Espinosa}\ \emph {et~al.}(2015)\citenamefont
	{Espinosa}, \citenamefont {Ryu}, \citenamefont {Marinov}, \citenamefont
	{Dumcenco}, \citenamefont {Kis},\ and\ \citenamefont
	{Garcia}}]{Espinosa:2015_Appl.Phys.Lett.}%
\BibitemOpen
\bibfield  {author} {\bibinfo {author} {\bibfnamefont {F.~M.}\ \bibnamefont
		{Espinosa}}, \bibinfo {author} {\bibfnamefont {Y.~K.}\ \bibnamefont {Ryu}},
	\bibinfo {author} {\bibfnamefont {K.}~\bibnamefont {Marinov}}, \bibinfo
	{author} {\bibfnamefont {D.}~\bibnamefont {Dumcenco}}, \bibinfo {author}
	{\bibfnamefont {A.}~\bibnamefont {Kis}}, \ and\ \bibinfo {author}
	{\bibfnamefont {R.}~\bibnamefont {Garcia}},\ }\href {\doibase
	10.1063/1.4914349} {\bibfield  {journal} {\bibinfo  {journal} {Appl. Phys.
			Lett.}\ }\textbf {\bibinfo {volume} {106}},\ \bibinfo {pages} {103503}
	(\bibinfo {year} {2015})}\BibitemShut {NoStop}%
\bibitem [{\citenamefont {Hancox}\ \emph {et~al.}(2010)\citenamefont {Hancox},
	\citenamefont {Sullivan}, \citenamefont {Chauhan}, \citenamefont {Beaumont},
	\citenamefont {Rochford}, \citenamefont {Hatton},\ and\ \citenamefont
	{Jones}}]{Hancox:2010_Org.Electron.}%
\BibitemOpen
\bibfield  {author} {\bibinfo {author} {\bibfnamefont {I.}~\bibnamefont
		{Hancox}}, \bibinfo {author} {\bibfnamefont {P.~J.}\ \bibnamefont
		{Sullivan}}, \bibinfo {author} {\bibfnamefont {K.~V.}\ \bibnamefont
		{Chauhan}}, \bibinfo {author} {\bibfnamefont {N.}~\bibnamefont {Beaumont}},
	\bibinfo {author} {\bibfnamefont {L.~A.}\ \bibnamefont {Rochford}}, \bibinfo
	{author} {\bibfnamefont {R.~A.}\ \bibnamefont {Hatton}}, \ and\ \bibinfo
	{author} {\bibfnamefont {T.~S.}\ \bibnamefont {Jones}},\ }\href {\doibase
	10.1016/j.orgel.2010.09.014} {\bibfield  {journal} {\bibinfo  {journal} {Org.
			Electron.}\ }\textbf {\bibinfo {volume} {11}},\ \bibinfo {pages} {2019}
	(\bibinfo {year} {2010})}\BibitemShut {NoStop}%
\end{thebibliography}

%


\end{document}